\documentclass[physrev,reprint,superscriptaddress,bibnotes,aps]{revtex4-2}  

\usepackage[T1]{fontenc}
\usepackage[australian]{babel}
\usepackage[a4paper,centering,hmargin=1.75cm,vmargin=2cm]{geometry} 

\usepackage{amsmath,amssymb,graphicx,bm,microtype} 
\usepackage[dvipsnames]{xcolor} 
\usepackage{booktabs} 
\usepackage[colorlinks,allcolors=blue!50!black]{hyperref} 
\usepackage[all]{hypcap} 
\usepackage{cleveref} 
\usepackage{orcidlink}
\newcommand{\orcid}[1]{\,\orcidlink{#1}}

\usepackage{braket} 
\usepackage{upgreek,textgreek} 
\usepackage{siunitx,textcomp} 
\DeclareMathSymbol{\ii}{\mathalpha}{letters}{"10} 
\DeclareMathSymbol{\jj}{\mathalpha}{letters}{"11} 
\renewcommand{\vec}[1]{\bm{\mathrm{#1}}}

\newcommand{\dg}{\dagger}
\newcommand{\D}{\mathrm{D}}
\newcommand{\A}{\mathrm{A}}

\begin{document}

\title{Unifying Collective Effects in Emission, Absorption, and Transfer}

\author{Adesh Kushwaha\orcid{0009-0008-2995-4568}}
\affiliation{School of Chemistry, University of Sydney, NSW 2006, Australia}

\author{Erik M. Gauger\orcid{0000-0003-1232-9885}}
\affiliation{SUPA, Institute of Photonics and Quantum Sciences, Heriot-Watt University, Edinburgh EH14 4AS, United Kingdom}

\author{Ivan Kassal\orcid{0000-0002-8376-0819}}
\email[Email: ]{ivan.kassal@sydney.edu.au}
\affiliation{School of Chemistry, University of Sydney, NSW 2006, Australia}


\begin{abstract}
    Collective effects, such as superradiance and subradiance are central to emerging quantum technologies—from sensing to energy storage—and play a important role in light-harvesting. These effects enhance or suppress rates of dynamic processes (absorption, emission, and transfer) due to the formation of symmetric or antisymmetric collective states. However, collective effects in different contexts---absorption, emission, and transfer---have often been defined disparately, especially across different communities, leading to results that are not immediately transferable between different contexts. Here, we describe all three types of collective effects using a common Dicke framework that resolves the apparent discrepancies between different approaches. It allows us to generalise previously known collective effects involving spins into new ones involving aggregates of harmonic oscillators or other degrees of freedom. It also explains how collective effects can be engineered to be robust against both disorder and noise, paving the way for more resilient quantum devices.
\end{abstract}

\maketitle

Collective effects arise when many particles behave in a way that cannot be understood by considering them as independent constituents. A broad class of collective effects arise in emission~\cite{Dicke1954}, absorption~\cite{Higgins2014}, and transfer~\cite{Strk1977,Lloyd2010}, where they modify rates of dynamical processes through coherent delocalisation among many particles. These types of collective effects have proposed applications in light harvesting~\cite{Liu2023}, ultra-narrow lasers~\cite{Meiser2009}, super-resolution imaging~\cite{Gorlach2024}, sensing~\cite{Zhu2024}, and quantum batteries~\cite{Andolina2019,Quach2022,Campaioli2024,Dias2024}. Collectively suppressed rates offer protection from radiative losses for quantum memories~\cite{Lvovsky2009} and ultra-narrow spectra for subradiance spectroscopy~\cite{McGuyer2014,Pasquiou2014,McDonald2017}. We will refer to collective effects that modify the rates of emission, absorption, and transfer as CEEAT.

The best-known CEEAT is superradiance (SR), where emitters collectively radiate light faster than they would individually~\cite{Dicke1954}. Collective rate enhancement also occurs in absorption and transfer. Superabsorption~(SA) is the time reversal of~SR, i.e., the collectively enhanced absorption of light~\cite{Higgins2014,Quach2022}. Supertransfer~(ST), by contrast, is the transfer of energy, charge, or other carriers enhanced by the collective behaviour of either donors or acceptors (or both)~\cite{Strk1977,Lloyd2010,Baghbanzadeh2016}. Because CEEAT is a dynamical category, it does not include related equilibrium effects such as superradiant phase transitions~\cite{Hepp1973_a,Wang1973, Kirton2019}.

Rather than enhancing, collective effects can also reduce rates of emission, absorption, and transfer, processes that are respectively referred to as subradiance~(SubR)~\cite{Freedhoff1967}, subabsorption~(SubA)~\cite{Gold2025}, and subtransfer~(SubT). CEEAT can produce completely ``dark'' states, which are decoupled from the environment or dynamical channels and thus lead to no dynamics. The decoupling makes such states immune to environmental noise, which can make them useful as decoherence-free subspaces~\cite{Lidar1998,Kwiat2000} in quantum information theory.

We use the word ``excitation'' to describe any entity involved in CEEAT. They can be atomic excitations, which cause emission in SR or are created in SA; excitons, transferred in light harvesting through generalised F\"orster resonance energy transfer (gFRET)~\cite{Sumi1998,Baghbanzadeh2016}; or charges, transferred through generalised Marcus theory~\cite{Taylor2018}.

Several CEEAT have been experimentally observed, but experimental challenges have left others undemonstrated or with few demonstrations. Of CEEAT enhancements, SR was observed in a gas in 1973~\cite{Skribanowitz1973}, while observations of SA are more recent~\cite{Yang2021,Quach2022}. Suppressed collective effects are experimentally more challenging because preparing and manipulating dark states is difficult, as they cannot be directly optically addressed; however, indications of SubR were reported in 1985~\cite{Pavolini1985} (with more direct experimental evidence in 2016~\cite{Guerin2016}), while SubA experiment is very recent~\cite{Gold2025}. Both ST and SubT have yet to be directly observed, although doing so should be possible using circuit quantum electrodynamics~\cite{Potonik2018,Kushwaha2025}. 
Given the broad range of effects, applications and physical systems, it is unsurprising that CEEAT have been treated in disparate ways in the literature, making ideas less transferable between different contexts. There are three aspects to this challenge. First, there are inconsistencies in how different collective effects are defined, especially in different research communities: some researchers, especially many physicists, typically expect rate enhancements to scale quadratically with the number of particles (as $N^2$)~\cite{Dicke1954,Gross1982,Cong2016,Nefedkin2017,Bojer2022,Sierra2022,Liedl2024,Holzinger2025}, others, especially chemists, accept linear scaling (as $N$)~\cite{Scully2009,Spano1991,Baghbanzadeh2016,Choudhary2019,Zhu2024_single}, and some define the phenomenon as any rate greater (or less) than that of a single emitter~\cite{Freedhoff1967,Stiesdal2020,Glicenstein2022,Pennetta2022}. Second, most CEEAT work has focused on spin systems, with much less work devoted to collective effects involving other degrees of freedom, such as harmonic oscillators (HOs). While SR for HOs has received some attention~\cite{Zakowicz1971,delanty2011superradianceharmonicoscillators,Delanty2012,Orell2022}, SA and ST in these systems remain unexplored. Third, the robustness of CEEAT to environmental effects has also been treated differently in different communities.

Here, we unify all CEEAT in a Dicke framework that resolves definitional disparities and allows us to describe new collective effects involving other degrees of freedom, including HOs. We show that CEEAT occur between a donor~($\mathrm{D}$) and an acceptor~($\mathrm{A}$) when the excitations are suitably delocalised on the donor aggregate, the acceptor aggregates, or both. A further shared feature is that all CEEAT can be made robust to disorder and noise using suitable intra-aggregate couplings.

 \begin{figure*}
         \centering
         \includegraphics{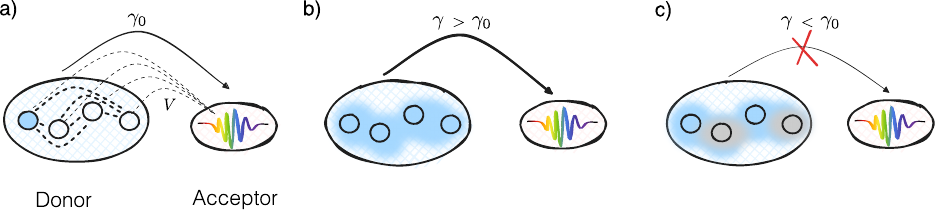}
            \caption{\textbf{Collectively enhanced and suppressed rate processes,} illustrated using super- and subradiance from a donor aggregate of four sites (blue) to an acceptor field.
            \textbf{(a)}~Emission from a localised donor excitation to the acceptor field occurs with rate $\gamma_0$. Dotted lines indicate couplings, including stronger intra-donor couplings (thicker) and weaker donor-acceptor couplings (thinner).
            \textbf{(b)}~Superradiance: symmetrically delocalised excitations can increase emission rates. 
            \textbf{(c)}~Subradiance: anti-symmetrically delocalised excitations can suppress the emission. }
         \label{fig:Collective_general}
\end{figure*}

\section{Overview}
\label{sec:Overview}
CEEAT changes the dynamical rate $\gamma$ of a transition between a donor~($\mathrm{D}$) and an acceptor~($\mathrm{A}$), as illustrated in~\cref{fig:Collective_general}. Either one or both of these may exist as an aggregate of components we will call ``sites''; typically, sites are taken to be two-level systems we will call ``spins'' (which may include atoms or molecules) or HOs (or bosons in general), but could be other degrees of freedom as well. Either the donor or the acceptor can also be a single entity with a continuous density of states, which we call a ``field'', which can include an electromagnetic field or the phonons in a solid. 

\begin{figure*}
        \centering
         \includegraphics{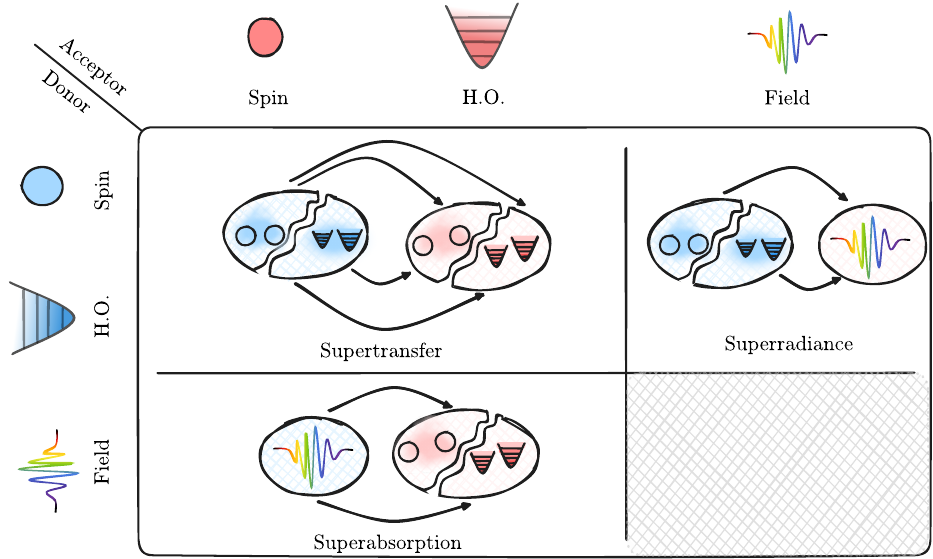}
            \caption{\textbf{Classification of collective enhancements in emission, absorption, and transfer.} The donor (blue) and the acceptor (red) may consist of spins, harmonic oscillators (HOs), or a field. Superradiance is the enhanced emission from spins or HOs into a field, while superabsorption is the inverse process of enhanced absorption from a field. Supertransfer is the enhanced transfer between one aggregate of spins or HOs and another. An analogous table could classify the suppressed collective effects into subradiance, subabsorption, and subtransfer.}
         \label{fig:Collective_table}
\end{figure*}

The different cases of CEEAT are categorised in~\cref{fig:Collective_table}, depending on the origin and target of the transition: emission into a field is SR, absorption from a field is SA, and transfer between aggregates is ST. In general, the CEEAT Hamiltonian is
\begin{equation}
\label{eqn:H_general}
    H=H_\D+H_\A+H_{\D\A},
\end{equation}
where $H_\D$ describes the donor, $H_\A$ the acceptor, and $H_{\D\A}$ the interaction between them.

CEEAT effects modify the rate $\gamma$ of excitation transfer from the donor to the acceptor. We consider the transfer rate from initial state $\ket{i}$ to final state $\ket{f}$, which can be calculated using Fermi's golden rule~\cite{Fox2006,May2011}, 
    \begin{equation}
        \label{eqn:Fermi_rate}
        \gamma_{i\to f}=\frac{2\pi}{\hbar}\left|\bra{f}H_{\D\A}\ket{i}\right|^2\nu(\omega_{fi}),
    \end{equation}
where $\nu(\omega_{fi})$ is the density of states near the energy difference $E_f-E_i$. For the golden rule to be valid, $\ket{i}$ and $\ket{f}$ must be eigenstates of the system, $H_{\D\A}$ must be weak compared to the relevant energy gaps, and there must be a continuum of final states to allow irreversible transitions. In SR and SA, this continuum is supplied intrinsically by the field, while in ST it necessarily originates from the aggregates' coupling to an environment. Additional broadening due to environmental noise or damping can further smooth the density of states, but is not required for SR and SA.

We then quantify CEEAT enhancements as the ratio $\gamma_{i\to f}/\gamma_0$, where $\gamma_0$ is the rate of normal transfer, i.e., when both $\ket{i}$ and $\ket{f}$ are localised on individual sites (or an incoherent mixture of them). 

The rates $\gamma_{i\to f}$ can be used to calculate other dynamical properties. The total rate of transfer from donor to acceptor depends on the populations $p_i$ of donor initial states, which may change over time as transitions take place, $\gamma_{\D\to \A}(t) = \sum_{i,f} p_i(t) \gamma_{i\to f}$. Similarly, the net excitation flux between the donor and the acceptor is given by the difference between the forward and backward transfers, $\gamma_\mathrm{net}(t) = \gamma_{\D\to \A}(t) - \gamma_{\A\to \D}(t)$. Finally, in SR, the rate is closely related to, and usually proportional to, the intensity of emission; whereas the rate indicates the speed of the emission of photons, the intensity is the speed of the emission of energy. However, for simplicity, we will write $\gamma$ instead of $\gamma_{i\to f}$ below.

CEEAT effects are due to the delocalisation of either $\ket{i}$ or $\ket{f}$ (or both). As we show below, the maximum rate enhancement occurs when the excitations are coherently delocalised with all sites in phase. Conversely, suppression can occur if out-of-phase delocalisation leads to destructive interference. If both the donor and the acceptor are delocalised, each can contribute a separate enhancement factor towards the rate.

\subsection{Example}
\label{sec:example}

To illustrate the main features of CEEAT, we take as an example SR from an aggregate of four spins, as shown in~\cref{fig:Collective_general}.

To show enhancement due to delocalisation, we consider an eigenstate where a single excitation is delocalised symmetrically over the four donor sites ($N_\D=4$) as in~\cref{fig:Collective_general}b,
    \begin{equation}
        \label{eqn:Delocalised_Donors_bright}
        \ket{D_+^{(1)}}=\frac{1}{\sqrt 4}(\ket{D_1}+\ket{D_2}+\ket{D_3}+\ket{D_4}),
    \end{equation}
where $\ket{D_j}$ describes an excitation on site $j$. Emission from $\ket{D_+^{(1)}}$ into the vacuum brings the donors to the ground state $\ket{D^{(0)}}$, with rate
\begin{equation}
\label{eqn:Classical_super}
\gamma \propto \left|\bra{D^{(0)}}H_\mathrm{DA}\ket{D_+^{(1)}}\right|^2 \quad\Rightarrow\quad\gamma/\gamma_0 = 4,
\end{equation}
assuming that $\bra{D^{(0)}}H_\mathrm{DA}\ket{D_j}$ is equal for all sites.
The enhancement arises because, in \cref{eqn:Classical_super}, the four amplitudes $\bra{D^{(0)}}H_\mathrm{DA}\ket{D_j}$ add in phase, resulting in constructive interference.
This is a special case of the general result, given below, that the enhancement for a single excitation is at most $\gamma/\gamma_0 = N_\D$. 

Even stronger emission, scaling as $N_\D^2$, is possible when there are more excitations in the system. In our example, two excitations in the state
    \begin{multline}
        \ket{D^{(2)}_+}=\frac{1}{\sqrt 6}(\ket{D_1}\ket{D_2}+\ket{D_1}\ket{D_3}+\ket{D_1}\ket{D_4} \\+\ket{D_2}\ket{D_3}+\ket{D_2}\ket{D_4}+\ket{D_3}\ket{D_4})      
    \end{multline}
yield the emission rate
\begin{equation}
    \label{eq:foursite-twoexciton}
    \gamma\propto \left|\bra{D_+^{(1)}}H_\mathrm{DA}\ket{D^{(2)}_+}\right |^2\quad\Rightarrow\quad\gamma/\gamma_0=6,
\end{equation}
which is the $N_\D=4$ case of the maximum enhancement factor of $N_\D(N_\D+2)/4$, derived below. The larger enhancement reflects the greater number of terms that interfere constructively in \cref{eq:foursite-twoexciton}. 

An anti-symmetric delocalised eigenstate shows suppression of emission. For the example in~\cref{fig:Collective_general}c, a single excitation in the state
    \begin{equation}
        \label{eqn:Delocalised_Donors_dark}
        \ket{D_-^{(1)}}=\frac{1}{\sqrt 4}(\ket{D_1}-\ket{D_2}+\ket{D_3}-\ket{D_4}),
    \end{equation}
leads to the emission rate 
\begin{equation}
    \label{eq:foursite-dark}
    \gamma\propto \left|\bra{D^{(0)}}H_\mathrm{DA}\ket{D_-^{(1)}}\right |^2 =0.
    \end{equation}
The suppression of the rate arises from the destructive interference of the terms in \cref{eq:foursite-dark}.

\begin{figure*}
    \centering
    \includegraphics[width=\linewidth]{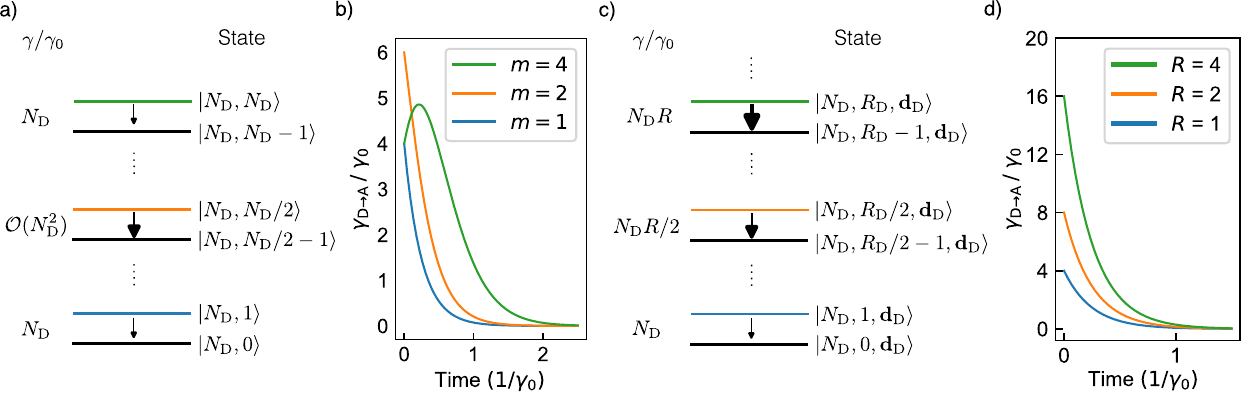}
    \caption{\textbf{Scaling of the rate $\gamma$ for Dicke states of spins and HOs.} 
    \textbf{(a)} Dicke states of an aggregate of $N$ spins are labelled $\ket{\ell,m}$; shown is only the Dicke ladder of the bright states with $\ell=N$ because the other states are dark. Going up the ladder, the enhancement increases from $\mathcal{O}(N)$ at $m=1$ to the maximum value of $\mathcal{O}(N^2)$ at half-filling ($m=N/2$) before decreasing back to $\mathcal{O}(N)$ for the fully excited aggregate. 
    \textbf{(b)} Time evolution of the rate out of a spin aggregate for different initial states. For initial states in the bottom half of the Dicke ladder, the rate decreases over time as excitations are lost. Initial states in the upper half, illustrated with $m=4$, show the characteristic bump in time, where the rate increases until the state passes through the middle of the ladder.
    \textbf{(c)} Dicke states of an aggregate of HOs are characterised by the occupation numbers of the collective modes, one bright with occupation number $R$ (the one shown) and $N-1$ dark ones, with occupation number vector $\mathbf{d}$. The bright mode emits with a rate that increases without bound as $N_\D R$.
    \textbf{(d)} Time evolution of the rate out of an HO aggregate for different initial states. Unlike the spin case, no bump is present for any initial $R$.
    }
    \label{fig:Dicke_ladder}
\end{figure*}

\subsection{Conflicting definitions}

A source of confusion in the literature is disparate criteria for using the terms ``super-'' and ``collective effects''. For concreteness, we continue using spin SR as our example, because the definitional ambiguities are most acute there; however, equivalent issues arise in discussions of SA and ST. There are broadly three definitions of collective effects based on the scaling of the rate enhancement: quadratic, linear, and general.

The narrowest, \textit{quadratic definition} is that a collective effect is one in which the rate scales as $\mathcal{O}(N^2)$ with the number of sites $N$~\cite{Dicke1954,Gross1982,Cong2016,Nefedkin2017,Bojer2022,Sierra2022,Liedl2024,Holzinger2025}. The argument for this approach is that $\mathcal{O}(N^2)$ scaling is truly ``collective'' because it requires multiple excitations, not just multiple sites. Specifically, the maximum rate occurs when the number of excitations is $N/2$. In a process obeying this definition, the statistics of the emitted photons display bunching, i.e., a second-order correlation function for which $g^{(2)}(0) > 1$~\cite{Mok2024}, compared to uncorrelated spontaneous emission with $g^{(2)}(0)=1$. Another feature commonly associated with the quadratic scaling is the presence of entanglement, as the dynamics proceeds through highly entangled Dicke states~\cite{Dicke1954}. However, entanglement is not necessary, because SR dynamics starting from a fully inverted state can be reproduced by fully separable mixtures of coherent spin states~\cite{Rosario2025}.

The \textit{linear definition} is that a collective effect is one where the rate scales (at least) as $\mathcal{O}(N)$~\cite{Scully2009,Spano1991,Baghbanzadeh2016,Choudhary2019,Zhu2024_single}. This approach is motivated by the desire to capture cooperative scenarios where typically only a single excitation of the $N$ sites is present~\cite{Spano1989}. Such conditions occur, for example, in single-photon SR~\cite{Scully2009} and in weakly excited light-harvesting complexes~\cite{Baghbanzadeh2016} or other molecular systems~\cite{Spano1989}. The linear enhancement in this regime has been described by Scully and Svidzinsky as ``the greatest radiation anomaly inherent in superradiance''~\cite{Scully2009}.

The broadest, \textit{general definition} captures any effect in which $\gamma$ is different from $\gamma_0$~\cite{Freedhoff1967,Stiesdal2020,Glicenstein2022,Pennetta2022}. It emerged out of the need to define SubR, which has no scaling laws with $N$~\cite{Freedhoff1967}. On this view, any emission faster than $\gamma_0$ is SR, and anything slower is SubR.

The definitions above are incomplete because they were conceived for spin CEEAT. When extended to other degrees of freedom, the general definition remains viable, while the quadratic and linear definitions can fail. In particular, for HOs the rate grows without bound with the number of excitations, making it impossible to speak of the scaling of the maximum rate~\cite{Zakowicz1971}.

The three definitions emphasise different features of CEEAT, and the choice among them depends on the problem in question. Here, we adopt the general definition, as it captures the widest range of effects and will allow us to construct the most inclusive framework for CEEAT. It can be applied both to suppressed effects and to any degree of freedom, including HOs. However, in the course of the discussion, we will point out effects that are characteristic of the more restrictive regimes.

\section{Spin collective effects in the Dicke limit}
\label{sec:spin}

We start by revisiting spin collective effects, which are well established~\cite{Dicke1954,Gross1982} and will form the base from which to generalise to new CEEAT effects in other degrees of freedom. In particular, we review the formalism of spin SR, whose mathematical structure offers a natural foundation for the description of SA and ST.

We first assume the Dicke limit~\cite{Dicke1954}: all sites are identical, the aggregate is much smaller than the wavelength of light involved, and intra-aggregate couplings (i.e., donor-donor and acceptor-acceptor interactions) are negligible, as are both disorder and noise. These assumptions ensure permutation symmetry of the Hamiltonian, allowing the spins to be mapped onto a single collective spin. In the Dicke limit, the aggregate eigenstates are delocalised even without intra-aggregate coupling because there is no disorder or noise, an assumption we will relax in \cref{sec:noise}.

\renewcommand{\arraystretch}{1.3}
\begin{table*}
    \centering
    \begin{tabular}{llllll}
        \toprule
        
             & Donor & Acceptor  & $\ket{i}=\ket{\cdot}_\D\ket{\cdot}_\A$& $\gamma/\gamma_0$  &Ref.\\
& &  & $\ket{f}=\ket{\cdot}_\D\ket{\cdot}_\A$& $\gamma_\mathrm{max}/\gamma_0$&\\
            \midrule
            \textbf{SR} & Spin & Field      & $\ket{N_\D,\ell_\D,m_\D}\ket{0}$& $(\ell_\D\!+\!m_\D\!-\!N_\D)(\ell_\D\!-\!m_\D\!+\!1)$&\cite{Dicke1954}\\
& &  & $\ket{N_\D,\ell_\D,m_\D\!-\!1}\ket{1}$& $\mathcal O (N_\D^2)$&\\
\cmidrule[0.1pt]{2-6}
                               & HO & Field      & $\ket{N_\D,R_\D,\vec{d}_\D}\ket{0}$    & ${N_\D R_\D}$     &\cite{Zakowicz1971}\\
& &  & $\ket{N_D,R_\D\!-\!1,\vec{d}_\D}\ket{1}$& $\infty$&\\
            \midrule
            \textbf{SA} & Field & Spin       & $\ket{1}\ket{N_\A, \ell_\A, m_\A}$& $(\ell_\A\!+\!m_\A\!-\!N_\A\!+\!1)(\ell_\A\!-\!m_\A)
$&\cite{Higgins2014}\\
& &  & $\ket{0}\ket{N_\A,\ell_\A,m_\A\!+\!1}$& $\mathcal O (N_\A^2)$&\\
\cmidrule[0.1pt]{2-6} 
            & Field & HO       & $\ket{1}\ket{N_\A,R_\A,\vec{d}_\A}$    & ${N_\A (R_\A\!+\!1)}$     &--\\
& &  & $\ket{0}\ket{N_\A,R_\A\!+\!1,\vec{d}_\A}$& $\infty$&\\
             \midrule
            \textbf{ST}  & Spin & Spin       & $\ket{N_\D,\ell_\D,m_\D}\ket{N_\A,\ell_\A,m_\A}$& $(\ell_\D\!+\!m_\D\!-\!N_\D)(\ell_\D\!-\!m_\D\!+\!1)(\ell_\A\!+\!m_\A\!-\!N_\A\!+\!1)(\ell_\A\!-\!m_\A)$&\cite{Lloyd2010}\\
& &  & $\ket{N_\D,\ell_\D,m_\D\!-\!1}\ket{N_\A,\ell_\A,m_\A\!+\!1}$& $\mathcal O (N_\D^2N_\A^2)$&\\
\cmidrule[0.1pt]{2-6} 
            &HO & HO        & $\ket{N_\D,R_\D,\vec d_\D}\ket{N_\A,R_\A,\vec d_\A}$     & $N_\D R_\D N_\A (R_\A\!+\!1)$&--\\
& &  & $\ket{N_\D,R_\D\!-\!1,\vec d_\D}\ket{N_\A,R_\A\!+\!1,\vec d_\A}$& $\infty$&\\
\cmidrule[0.1pt]{2-6} 
             & Spin & HO        & $\ket{N_\D,\ell_\D,m_\D}\ket{N_\A,R_\A,\vec d_\A}$& $(\ell_\D\!+\!m_\D\!-\!N_\D)(\ell_\D\!-\!m_\D\!+\!1){N_\A (R_\A\!+\!1)}$&--\\
& &  & $\ket{N_\D,\ell_\D,m_\D\!-\!1}\ket{N_\A,R_\A\!+\!1,\vec d_\A}$& $\infty$&\\
\cmidrule[0.1pt]{2-6} 
             & HO & Spin        & $\ket{N_\D,R_\D,\vec d_\D}\ket{N_\A,\ell_\A,m_\A}$& ${N_\D R_\D}(\ell_\A\!+\!m_\A\!-\!N_\A\!+\!1)(\ell_\A\!-\!m_\A)$&--\\
& &  & $\ket{N_\D,R_\D\!+\!1,\vec d_\D}\ket{N_\A,\ell_\A,m_\A\!-\!1}$& $\infty$&\\       
\bottomrule
    \end{tabular}
    \caption{\textbf{Scaling in enhanced collective effects.} The rate enhancement $\gamma/\gamma_0$ is given for all CEEAT effects, along with its maximum value $\gamma_\mathrm{max}/\gamma_0$. The rates are given for initial and final Dicke states. Spin states are in the form $\ket{N,\ell,m}$, denoting an aggregate of $N$ spins with total spin quantum number $\ell'=\ell-N/2$ and with $m$ excitations; the maximum values $\gamma_\mathrm{max}/\gamma_0$ are obtained for bright states ($\ell = N$) at half-filling ($m=\ell/2$). HO states are in the form $\ket{N,R,\vec d}$, which describes $N$ HOs with $R$ excitations in the symmetric bright mode and the list $\vec{d}$ of excitations in the dark modes; rates involving HOs have no maximum because $R$ is unbounded. References refer to prior derivations of the rates, where available.}
    \label{tab:Collective_effect}
\end{table*}

\textbf{Superradiance (SR).} SR from a spin aggregate (donor) into a field (acceptor) is described by the Hamiltonian in \cref{eqn:H_general} with
    \begin{subequations}
    \label{eqn:H_SR}
        \begin{align}
        H_\D^\mathrm{spin}&=\sum_{i=1}^{N_\D} \omega_\mathrm{D} \sigma_i^z,\quad
        H_\A^\mathrm{field}=\sum_{q}\omega_qb_q^\dagger b_q^{\vphantom{\dagger}}, \label{eqn:H_field}\\
        H_{\D\A}&= -i\sum_{q}\sum_{i=1}^{N_\D}\vec{\mu}_\D\cdot \vec{\epsilon}_qg(\sigma_i^+b_q+\sigma_i^-b_q^\dg). \label{eqn:H_int}
        \end{align}
        \end{subequations}
$H_\D^{\mathrm{spin}}$ describes an aggregate of $N_\D$ identical, uncoupled spins with transition frequency $\omega_{\mathrm{D}}$, where $\sigma_i^z$ is the Pauli $z$ operator for spin $i$. We assume $\hbar = 1$ throughout. The field is modelled as a continuum of modes with frequencies $\omega_q$ and annihilation operators $b_q$.
The spin-field interaction $H_{\mathrm{DA}}$ is a first-order coupling of the spin transition dipole moment $\vec\mu_{\mathrm{D}}$ to the field modes with interaction strength $g$, where $\vec\epsilon_q$ is the polarisation of mode $q$. Because we are in the weak-coupling regime, the rotating-wave approximation is valid, and the counter-rotating terms $\sigma^+b^\dg$ and $\sigma^-b$ are neglected.

Because \cref{eqn:H_SR} is permutationally symmetric, it can be solved using collective spin operators $J^+=\sum_i \sigma^+_i$ and $J^z=\sum_i \sigma^z_i$. In terms of these operators, $H$ is called the Dicke Hamiltonian, 
    \begin{multline}
            H_\mathrm{SR}^\mathrm{Dicke}=   \omega_\mathrm{D} J^z+\sum_{{q} }   \omega_q{b}_{{q} }^{\dagger} {b}_{{q} } {} \\
            -i\sum_{{q} } \vec{\mu}_\mathrm{D} \cdot \vec{\epsilon}_qg\left(J^{+}b_q+ J^{-}b_q^\dg\right). 
    \end{multline}
The eigenstates of $H_\mathrm{SR}^\mathrm{Dicke}$ are called Dicke states and are described by the eigenvalues $\ell_\D'(\ell_\D'+1)$ of $J^2=J_x^2+J_y^2+J_z^2$ and $m_\D'$ of $J_z$, where $J_{x,y}=\sum_i\sigma_i^{x,y}$~\cite{Sakurai2020-di}. Because of spin-addition algebra, these quantum numbers obey the inequality $N_\D/2 \geq \ell_\D'\geq |m_\D'| \geq 0$. For later consistency with HO SR, we adopt the relabelling
\begin{equation}
    \ket{N_\D, \ell_\D, m_\D} = \ket{N_\D, \ell_\D' + N_\D/2,\, m_\D' + N_\D/2},
\end{equation}
with $N_\D \geq \ell_\D \geq N_\D/2$ and $N_\D - \ell_\D \leq m_\D \leq \ell_\D$.
For example, the illustrative states in \cref{sec:example} can be expressed in this notation as
$\ket{D_+^{(2)}}=\ket{4,4,2}$, $\ket{D_+^{(1)}}=\ket{4,4,1}$, and $\ket{D_-^{(1)}}=\ket{4,3,1}$.

The bright states of the Dicke Hamiltonian are the fully symmetric states $\ket{N_\D, N_\D,m}$ (i.e., $\ell_\D=N_\D$), which can be represented using the Dicke ladder (\cref{fig:Dicke_ladder}a). They can be found by sequentially applying $J^+$ to the ground state~\cite{Sakurai2020-di}. All other states are dark.
In general, the rate enhancement when emitting an excitation ($m_\D \to m_\D-1$) is~\cite{Dicke1954}
\begin{multline}
\label{eqn:Dicke_rate}
    \gamma/\gamma_0  = |\left\langle N_\D, \ell_\D, m_\D-1\left|  J^- \right| N_\D,\ell_\D, m_\D \right\rangle|^2 \\ =(\ell_\D+m_\D-N_\D)(\ell_\D-m_\D+1).
\end{multline}
The rate is maximised when the spin system is half-filled ($m_\D=N_\D/2$), giving
\begin{equation}
    \gamma_\mathrm{max}/\gamma_0 = \frac{N_\D}{2}\left(\frac{N_\D}{2}+1\right)  = \mathcal{O}(N_\D^2),
\end{equation}
as required by the quadratic definition of CEEAT. For a single delocalised excitation ($m_\D=1$),  $\gamma_\mathrm{max}=N_\D\gamma_0$.

The dynamics of spin SR exhibits a characteristic superradiant bump in the temporal evolution of $\gamma(t)$, see \cref{fig:Dicke_ladder}b. When the system starts from the fully excited state ($m_\D=N_D$), the rate initially increases as the system descends the Dicke ladder, meaning that there is a delay before it reaches the maximum rate at half filling, after which it decreases again as the system continues to de-excite~\cite{Gross1982}.

\textbf{Superabsorption (SA).}  SA is the enhanced absorption from a field (donor) due to the delocalisation of an aggregate (acceptor). The Hamiltonian is the same as for SR, except that the roles of the donor and acceptor are reversed~\cite{Higgins2014,Yang2021}, i.e., the donor is described by $H_\D^\mathrm{field}$ and the acceptor by $H_\A^\mathrm{spin}$.
    
By time-reversal symmetry, the rate of absorption by a spin aggregate (taking $m_\A$ to $m_\A+1$) is the same as the corresponding emission rate in SR,
    \begin{multline}
        \gamma/\gamma_0  = |\left\langle N_\A, \ell_\A, m_\A+1\left|  J^+ \right| N_\A,\ell_\A, m_\A\right\rangle|^2  \\
         =(\ell_\A+m_\A-N_\A+1)(\ell_\A-m_\A).
    \end{multline}
The absorption rate from the ground state to the first delocalised excitation scales as $N_\A$. This enhancement increases further up the Dicke ladder until absorption starting from a half-filled Dicke state ($m_\A = N_\A/2$), which scales as $\mathcal{O}(N_\A^2)$. While it may seem counterintuitive that a half-filled state is easier to excite than a fully unexcited one, the higher rate for the half-filled state arises from the constructive interference of a larger number of transition amplitudes near half filling.

SA dynamics qualitatively changes as the excitation level approaches the half-filled Dicke state ($m_\A = N_\A/2$). At that point, SR emerges as a significant channel, competing with and potentially dominating SA. This competition can lead to transient behaviour where net absorption first accelerates and then slows down or even reverses. At constant optical driving, the steady state depends on the balance between the driving and the emission. Without additional control, emission dominates over absorption, leading to relaxation towards lower excitation. To sustain SA, emission must be inhibited, for instance, by engineering the electromagnetic environment to suppress emission channels~\cite{Higgins2014}.

\textbf{Supertransfer (ST).} 
ST is usually formulated in the single-excitation limit, as is appropriate for most molecular systems, in which case it is described using gFRET~\cite{Sumi1998,Baghbanzadeh2016}. However, parallels with multi-excitation SR show that ST can lead to as high as quartic scaling of~$\gamma$ with system size~\cite{Lloyd2010}.

ST between two spin aggregates is described by
        \begin{subequations}
    \label{eqn:H_ST}
        \begin{align}        
        H_\D^\mathrm{spin}&=\sum_{i=1}^{N_\D} \omega_\mathrm{D} \sigma_i^z,\quad
        H_\A^\mathrm{spin}=\sum_{j=1}^{N_\A} \omega_\mathrm{A} \sigma_j^z, \\
        H_{\D\A}&= \sum_{i=1}^{N_\D}\sum_{j=1}^{N_\A}V_\mathrm{DA}\left (\sigma_i^+\sigma_j^-+\sigma_i^-\sigma_j^+ \right ).
        \end{align}
        \end{subequations}
As in SR, we define collective operators $J_\D$ and $J_\A$, giving the Dicke Hamiltonian for spin ST,
    \begin{equation}
        H^\mathrm{Dicke}_\mathrm{ST}= {\omega_\D}J_\D^z+{\omega_\A} J^z_\A+
            V_\mathrm{DA}  \left ( J_\D^{+} J^{-}_\A+J^{-}_\D J^{+}_\A \right ).
    \end{equation}
    
A new feature of ST, compared to SR and SA, is that both donors and acceptors can be delocalised, giving a higher rate enhancement overall~\cite{Lloyd2010},
    \begin{align}
    \gamma/\gamma_0  
    & =\! |\langle N_\D,\ell_\D, m_\D\!-\!1 ; N_\A, \ell_\A, m_\A\!+\!1| {J}^{-}_\D {J}^{+}_\A \nonumber\\ 
    &\qquad\qquad|N_\D, \ell_\D, m_\D; N_\A,\ell_\A, m_\A\rangle|^2  \\
    & = (\ell_\D+m_\D-N_\D)(\ell_\D-m_\D+1)\times \nonumber\\
    &\qquad\qquad (\ell_\A+m_\A-N_\A+1)(\ell_\A-m_\A).
    \label{eqn:gamma_ST}
    \end{align}
In particular, the enhancement is a product of independent enhancement factors from the donor and the acceptor. Specifically, it is equal to the product of the enhancements for donor SR and acceptor SA.

ST provides the steepest scaling in collective enhancement of all spin CEEAT effects. 
The maximum ST rate occurs when both the donor and the acceptor are half-filled, $m_\D= N_\D/2,m_\A= N_\A/2$, when the maximum rate enhancement is $\gamma_\mathrm{max}/\gamma_0 =\mathcal{O}(N_\D^2N_\A^2)$~\cite{Lloyd2010}. In the specific case of equal-sized donor and acceptor, the enhancement can be quartic, $\mathcal{O}(N^4)$.

As mentioned in~\cref{sec:Overview}, the net rate of transfer between donor and acceptor, $\gamma_\mathrm{net}$, is reduced by the reverse transfer from acceptor to donor. An extreme case occurs if the donor and the acceptor have equal site energies, sizes, and excitation populations. In that case, the forward and backward rates are equal and $\gamma_\mathrm{net}=0$. Previous analysis predicted that $\gamma_\mathrm{net}$ can scale at most as $\mathcal{O}(N^3)$~\cite{Lloyd2010}, based on the assumption of symmetric transfer rates between donor and acceptor, in which case the quartic term $N_\mathrm{D}^2 N_\mathrm{A}^2 \gamma_0$ is equal for forward and backward transfers and cancels out. However, the quartic dependence is restored in $\gamma_\mathrm{net}$ when a significant energetic offset exists between donor and acceptor. In such cases, the density of states $\nu$ provided by the thermal environment strongly favours the downhill (donor-acceptor) transfer, while the reverse (uphill) process is heavily suppressed.

ST dynamics shows features common to both SR and SA because the rate enhancement is a product, \cref{eqn:gamma_ST}, of an SR-like contribution from the donor and an SA-like contribution from the acceptor. Both contributions follow the Dicke-ladder dynamics in~\cref{fig:Dicke_ladder}a,b; i.e., starting from a fully excited donor (or an empty acceptor), the rate increases toward half-filling and then decreases again as the donor moves further down (or the acceptor further up) the Dicke ladder. When the donor and acceptor are of comparable size ($N_\A \approx N_\D$), these two maxima occur at nearly the same stage of the dynamics. Therefore, the transfer rate rises faster and reaches a higher maximum ($\mathcal{O}(N_\D^4)$) at half-filling. When the acceptor is larger than the donor ($N_\A > N_\D$), its peak occurs later; in this case, after the donor has passed its peak and its contribution begins to fall, the acceptor’s contribution is still increasing toward its own maximum. This misalignment leads to a temporally broader bump in the transfer rate. The similarity between ST and SR bumps is clearest when $N_\A \gg N_\D$, because then the acceptor effectively mimics a field.

\section{Harmonic-oscillator collective effects in the Dicke limit}
\label{sec:ho}
The formalism reviewed above for spin collective effects provides a foundation for extending the description to non-spin degrees of freedom, which remain much less explored in the context of CEEAT. As a first example, we review the known HO SR case and then extend it to the new cases of SA and ST.

A crucial difference that arises in HOs, compared to spins, is that HOs can have more than one excitation per site, which leads to differences in their collective behaviour. Previous work on HO CEEAT has been restricted to HO SR~\cite{Zakowicz1971,delanty2011superradianceharmonicoscillators,Orell2022} and highlighted the main difference between spin and HO SR, i.e., that the rate enhancement is unbounded for the latter. Here, we extend the definitions of SR, SA, and ST to comprehensively describe CEEAT in aggregates involving HO modes, as illustrated in~\cref{fig:Collective_table}. As with spins, we begin in the Dicke limit, a restriction we will later relax.

\textbf{Superradiance.} SR occurs in delocalised aggregates of radiating HOs~\cite{Zakowicz1971}, whose Hamiltonian is
        \begin{subequations}
\label{eqn:H_SR_mode}
    \begin{align}        
    H_\D^\mathrm{HO} &=\sum_{i=1}^{N_\D} \omega_\mathrm{D}  a_i^\dagger a_i, \quad
    H_\A^\mathrm{field}=\sum_{q}\omega_qb_q^\dagger b_q^{{\dagger}}, \\
    H_{\D\A}&= -i\sum_q\sum_{i=1}^{N_\D} g_q(a_i^\dg b_q+a_ib_q^\dg ),
    \end{align}
    \end{subequations}
where $H_\D^\mathrm{HO}$ describes an aggregate of HO modes, each with frequency $\omega_\D$. The coupling to the field $H_{\mathrm{\D\A}}$ is assumed to be weak and is therefore expressed to first order in the coupling $g_q$.

As in spin SR, the emission rate can be calculated using~\cref{eqn:Fermi_rate} and collective operators~\cite{Zakowicz1971,delanty2011superradianceharmonicoscillators, Orell2022} that are a unitary transformation of the local operators $a_j$, 
\begin{equation}
    c_k=\frac{1}{\sqrt{ N_\D}}\sum_{j=1}^{N_\D}\exp{\left ( \frac{2\pi i}{N_\D}jk\right )a_j},
\end{equation}
for $k\in\{1,\ldots,N_\D\}$. The Dicke Hamiltonian is
\begin{multline}
    H^\mathrm{DickeHO}_\mathrm{ST}= \omega_\D\sum_{k=1}^{N_\D}c_k^\dg c_k+\sum_{q}\omega_qb_q^\dagger b_q \\   
    -i \sum_q ( c_{N_\D}^\dg b_q+c_{N_\D}b_q^\dg).
    \label{eqn:H_SR_Dicke_HO}
\end{multline}
Only the mode corresponding to $c_{N_\D}$ is coupled to the field and emits radiation; this sole bright state is a superposition with all modes in phase~\cite{Orell2022}, $c_{N_\D}=\sum_i a_i/\sqrt{{N_\D}}$. To distinguish it from the other modes, we label its number of excitations as $R_\D$. The remaining $N_\D-1$ modes do not couple to the field and are thus dark; we denote the number of excitations in them by the vector $\vec d_\D = (d_1,d_2,\ldots,d_{{N_\D}-1})$.

The overall state of the HO aggregate is determined by $R_\D$ and $\vec d_\D$ and can be expressed as excitations of the ground state $\ket{0}$ as
    \begin{equation}
        \ket{{N_\D},R_\D,\vec d_\D}=\frac{({c}_N^{\dagger})^{R_\D}}{\sqrt{R_\D!}} \prod_{i=1}^{{N_\D}-1} \frac{({c}_i^{\dagger})^{d_i}}{\sqrt{d_{i}!}}\ket{0}_\mathrm{L}^{
        \otimes {N_\D}}.
    \end{equation}
For example, for an aggregate with two HOs, the lowest-occupancy states in the Dicke basis can be re-expressed in the local Fock basis $\ket{n_1n_2}_\mathrm{L}$ as follows:
\begin{subequations}
     \begin{align}
     \ket{2,2,\{0\}}&=(\ket{20}_\mathrm{L}+\ket{02}_\mathrm{L}+\sqrt{2}\ket{11}_\mathrm{L})/2,\\
        \ket{2,1,\{0\}}&=(\ket{10}_\mathrm{L}+\ket{01}_\mathrm{L})/\sqrt{2},\\
        \ket{2,0,\{1\}}&=(\ket{10}_\mathrm{L}-\ket{01}_\mathrm{L})/\sqrt{2},\\
        \ket{2,0,\{0\}}&=\ket{00}_\mathrm{L}.
    \end{align}
\end{subequations}
Here, the states $\ket{2,1,\{0\}}$ and $\ket{2,2,\{0\}}$ lead to enhanced emission rates because their symmetric superpositions cause constructive interference of transition amplitudes, whereas $\ket{2,0,\{1\}}$ results in a suppressed rate due to its antisymmetric character. 

In general, the rate enhancement for emission, when moving down the Dicke ladder from $R_\D$ bright excitations to $R_\D-1$, is
\begin{multline}
\label{eqn:gamma_SR_HO}
    \gamma/\gamma_0 = |\langle {N_\D},R_\D-1,\vec d_\D|{c}_{N_\D} | {N_\D},R_\D,\vec d_\D\rangle |^2 \\ = {N_\D} R_\D.
\end{multline}
Therefore, emission is enhanced by a factor of $N_\D R_\D$~\cite{delanty2011superradianceharmonicoscillators} compared to emission from an individual HO. Because there is no upper bound on $R_\D$, this enhancement can be arbitrarily large, unlike in spin SR.

A distinguishing feature of HO SR is that the peak rate is not temporally delayed, i.e., there is no bump in emission as there is in spin SR~\cite{Orell2022}, see~\cref{fig:Dicke_ladder}d. In HO aggregates, higher excitations always lead to higher rates (in proportion to $R$), meaning that emission slows as excitations are emitted.

\textbf{Superabsorption.} We can now extend SA from the spin case to the previously undescribed enhanced absorption by an aggregate of HOs. The Hamiltonian for HO SA is identical to that of SR (\cref{eqn:H_SR_Dicke_HO}), but with the role of donor and acceptor reversed. The absorption rate is determined by calculating $\gamma$ from \cref{eqn:Fermi_rate}, which involves the absorption of a photon to increase the number of bright acceptor excitations from $R_\A$ to $R_\A+1$,
\begin{multline}
\label{eqn:rate_HO_SA}
    \gamma/\gamma_0  = |\langle {N_\A},R_\A+1,\vec d_\A|{c}_{N_\A}^{\dagger} | {N_\A},R_\A,\vec d_\A\rangle |^2 \\ ={N_\A} (R_\A+1).
\end{multline}
As with SR, the absorption rate in SA is unbounded, a feature that is only possible for unbounded degrees of freedom such as HOs.

\textbf{Supertransfer.} We also extend the concept of ST to include enhanced excitation transfer between aggregates of HOs. The Hamiltonian is
\begin{multline}
    H_\mathrm{ST}^\mathrm{DickeHO}=\sum_{i=1}^{{N_\D}}\omega_\mathrm{D} {a}_i^{\dagger} {a}_i+\sum_{j=1}^{N_\A} \omega_\mathrm{A} {b}_j^{\dagger} {b}_j\\ + \sum_{ i=1}^{N_\D}\sum_{j=1}^{N_\A}V_{\mathrm{DA}}({a}_i^{\dagger} {b}_j+{a}_i {b}_j^{\dagger}).
\end{multline}
This system can be analysed using the Dicke collective operators above. Because there is now delocalisation both among donors and among acceptors, the transfer enhancement scales as
\begin{align}
    \gamma/\gamma_0 &= |\bra{N_\D,R_\D-1,\vec d_\D;N_\A,R_\A+1,\vec d_\A} {c}_{{N_\D}} {c}^{\dg}_{{N_\A}} \nonumber \\
    &\quad\quad\quad \ket{N_\D,R_\D,\vec d_\D;N_\A,R_\A,\vec d_\A}|^2 \nonumber \\
    &= N_\D N_\A R_\D(R_\A+1).
\end{align}
As in the spin case, ST among HOs exhibits the most rapid rate scaling among CEEAT effects. Since $R_\D$ and $R_\A$ are unbounded, the corresponding enhancement in the rate is also unbounded.

\begin{figure*}
     \centering
     \includegraphics[width=1\linewidth]{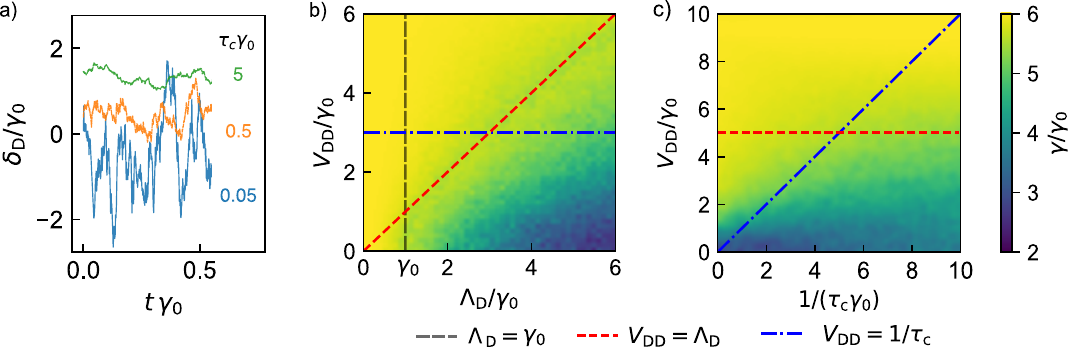}
        \caption{\textbf{Robust CEEAT due to intra-aggregate couplings} in a four-spin donor aggregate ($N_{\mathrm{D}}=4$), initialised in the half-filled Dicke state. 
        \textbf{(a)}~Samples of Ornstein--Uhlenbeck noise with strength $\Lambda=\gamma_0$ and a range of correlation times.
        \textbf{(b,c)}~The rate enhancement $\gamma/\gamma_0$, obtained from the emission $\gamma_0\langle J^{+} J^{-} \rangle$ averaged over evolution up to $t=1/\gamma_0$. Maximal enhancement $\gamma_{\max}/\gamma_0 = 6$ is achieved when the intra-aggregate coupling $V_{\D\D}$ is large compared to both the noise strength $\Lambda_\D$ and the inverse correlation time $\tau_\mathrm{c}$, i.e., $V_{\D\D} \gg \Lambda_\D,\tau_\mathrm{c}^{-1}$.
        (b)~Rate enhancement as a function of $V_{\mathrm{DD}}$ and $\Lambda_{\mathrm{D}}$ at fixed $\tau_\mathrm{c}=0.33\,\gamma_0^{-1}$. An additional way to preserve CEEAT occurs when $\Lambda_{\mathrm D} \ll \gamma_0$, even for no coupling.
        (c)~Rate enhancement as a function of $V_{\mathrm{DD}}$ and $\tau_\mathrm{c}^{-1}$ at fixed $\Lambda_{\mathrm D}=5\gamma_0$.
        }
    \label{fig:Noise_effect}
\end{figure*}

\section{Mixed Collective Effects}

Our approach to CEEAT can be readily extended to mixed ST systems, in which one aggregate is composed of spins and the other of HOs. In the mixed case, the spin Dicke operators of \cref{sec:spin} can be used to describe the spin aggregate and the HO Dicke operators of \cref{sec:ho} the HO aggregate. The results of these calculations are given in \cref{tab:Collective_effect}, showing a pattern where the enhancement is a product of a factor due to delocalisation among the spins and a factor due to delocalisation among the HOs. In all mixed cases, the maximum ST rates remain unbounded due to the possibility of unlimited bright excitations in the HO aggregate.

As an example, ST between a donor spin aggregate and an acceptor HO aggregate is described by $H_\D^\mathrm{spin}$, $H_\A^\mathrm{HO}$, and the coupling
\begin{align}
\label{eqn:H_ST_hybrid}
H_{\D\A}= \sum_{i=1}^{N_\D}\sum_{j=1}^{N_\A}V_\mathrm{DA}\big(\sigma_i^+ b_j+\sigma_i^- b_j^{\dagger} \big).
\end{align}
As in \cref{sec:spin,sec:ho}, we define a collective spin operator $J_\D$ for the donor and a collective bosonic mode $c_{N_\A}$ for the acceptor, giving
\begin{equation}
H^\mathrm{mix}= {\omega_\D}J_\D^z+\sum_{j=1}^{N_\A}{\omega_\A} b_j^{\dagger} b_j+
V_\mathrm{DA} \big( J_\D^{+} c_{N_\A}+J^{-}_\D c_{N_\A}^{\dagger} \big).
\end{equation}
The transfer enhancement for $H^\mathrm{mix}$ scales as 
\begin{align}
{\gamma}/{\gamma_0}
&=\big|\bra{N_\D,\ell_\D, m_\D\!-\!1}\bra{ N_\A, R_\A\!+\!1,\vec d_\A}
J^{-}_\D c^{\dagger}_{N_\A} \nonumber\\
&\qquad\qquad\ket{N_\D, \ell_\D, m_\D}\ket{ N_\A,R_\A,\vec d_\A}\big|^2 \nonumber\\
&= (\ell_\D+m_\D-N_\D)(\ell_\D-m_\D+1)N_\A (R_\A+1).
\end{align}
Here, the factors with the subscript D are exactly the Dicke enhancement for spin SR (\cref{eqn:Dicke_rate}), while the those with the subscript A equal the enhancement for HO SA (\cref{eqn:rate_HO_SA}).

\section{Beyond Spins and Harmonic Oscillators}

CEEAT can also be extended to systems with other degrees of freedom, where the intuitions derived from the spin and HO cases remain valuable. For example, core CEEAT features remain present in prior descriptions of SR from aggregates of qudits (multi-level systems)~\cite{Crubellier1980,Sutherland2017} and anharmonic oscillators~\cite{Orell2022}, all of which feature delocalisation that modifies transfer rates. 

A representative example is SR in anharmonic oscillators~\cite{Fisher1989,Roushan2017,Orell2022}, where each donor site has a finite on-site anharmonicity $U$ that penalises multiple excitations,
\begin{multline}
    H_\mathrm{SR}^{\mathrm{anh}} = \sum_{i=1}^{N_\mathrm{D}}
    \big(\omega_\mathrm{D}\, a_i^\dagger a_i
    +\tfrac{1}{2}U\, a_i^\dagger a_i(a_i^\dagger a_i - 1)\big)
    \\
    +\sum_{j=1}^{N_\mathrm{A}}\omega_\mathrm{A}\, b_j^\dagger b_j
    +\sum_{i,j}V_{\mathrm{DA}}
    \big(a_i^\dagger b_j + a_i b_j^\dagger\big).
\end{multline}
As $U$ decreases, the donor sites become increasingly harmonic; in the opposite limit, they becomes more spin-like. Consequently, the collective eigenstates involved in SR are no longer purely HO Dicke states but instead contain both spin-like single-occupancy components and HO-like multi-occupancy components~\cite{Fisher1989,Roushan2017,Orell2022}. For small $U$, multi-occupancy configurations are not penalised and SR approaches the unbounded HO scaling, \cref{eqn:gamma_SR_HO}. As $U$ increases, these multi-excitation components are progressively suppressed, the states become increasingly spin-like, and the transfer rate approaches the spin Dicke limit, \cref{eqn:Dicke_rate}. 

Within our framework, anharmonic SR effects generalise naturally to superabsorption (SA) and supertransfer (ST), including mixed ST where donors and acceptors differ. For anharmonic-oscillator ensembles, both SA and ST rates would fall between the spin and harmonic-oscillator limits, remaining unbounded with increasing excitation number while still staying below the harmonic oscillator rate for the same excitation number.

\begin{figure}
    \centering
    \includegraphics[width=0.85\linewidth]{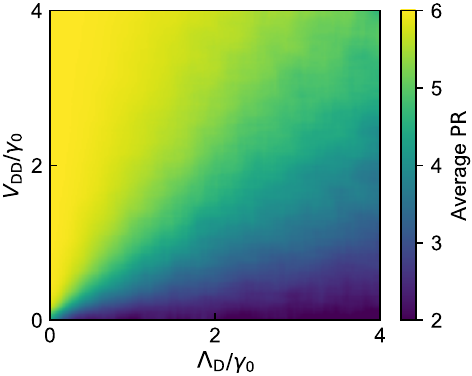}
    \caption{\textbf{Robust delocalisation due to intra-aggregate couplings} in a disordered four-spin donor aggregate ($N_\D=4$). The average participation ratio (PR) of the two-particle eigenstates as a function of intra-aggregate coupling $V_{\D\D}$ and the variance $\Lambda$ of the static energetic disorder. A higher PR, up to a maximum of $\mathrm{PR_{max}}=6$, indicates greater delocalisation of the states, and is obtained when $\Lambda_\D \ll V_{\D\D}$. }
    \label{fig:PR_noise}
\end{figure}

\section{Coupling-Induced Robustness against Disorder and Noise}
\label{sec:noise}

In realistic systems, disorder and noise break the permutation symmetry of the Dicke model and tend to localise excitations on individual sites, suppressing CEEAT~\cite{Celardo2014_static,Celardo2014_dynamic,Chen2022}. However, intra-aggregate couplings can counteract this localising tendency, thereby extending collective effects beyond the Dicke limit.

Using couplings to counteract the localising effects of disorder and noise has been explored in several CEEAT contexts, e.g., in SR of quantum dots~\cite{Miftasani2016, Blach2022,Luo2025}, molecular aggregates~\cite{Spano1989,Potma1998} and the 1D Anderson model~\cite{Celardo2013}.
When such couplings are present, they can give rise to collective behaviour reminiscent of the Dicke case, while also enhancing robustness against environmental perturbations~\cite{Celardo2014_static,Celardo2014_dynamic}. In biological light-harvesting systems, such as the LH2 complex or purple bacteria, intra-aggregate couplings can lead to excitonic delocalisation that extends over small clusters of chromophores and remains surprisingly robust in the presence of disorder and noise~\cite{Smyth2015,Baghbanzadeh2016,Taylor2018}. Similar phenomena are observed in organic photovoltaics, where even small amounts of delocalisation can significantly improve device performance~\cite{Balzer2022,Balzer2023,Balzer2024}. A related idea in polaritonic systems is to enhance delocalisation in disordered aggregates through stronger couplings mediated by a cavity~\cite{Ribeiro2018,Engelhardt2023,Sanchez2024,Liu2025}. 

CEEAT beyond the Dicke limit involves three parameters that are all zero in the Dicke model~\cite{Gross1982,May2011}: disorder, noise, and intra-aggregate couplings. 

Disorder and noise are the two extreme limits of sources of symmetry breaking. Disorder refers to random variations in site energies or couplings fixed by the material structure. Site-energy disorder is usually modelled by replacing the constant site energies $\omega$ (in, e.g., \cref{eqn:H_SR}) with $\omega + \delta_i$, where $\delta_i$ is a site-dependent energy shift. Couplings (either within or between aggregates) can similarly be disordered. By contrast, noise denotes time-dependent fluctuations, caused by a coupling to an environment, that cause dephasing and energy relaxation. Although there are many ways to describe noise, here we assume that it acts as local site-energy fluctuations, i.e., the $\delta_i(t)$ become time dependent.

We can model disorder, noise, and intermediate processes within a single framework by assuming the site-energy fluctuations are independent, identically distributed Ornstein–Uhlenbeck (OU) stochastic processes~\cite{Uhlenbeck1930,Gardiner1985},
\begin{equation}
    d\delta_i(t)
      = -\frac{1}{\tau_c}\,\delta_i(t)\,dt 
        + \sqrt\frac{2\Lambda}{\tau_c}\, dW(t),
\end{equation}
where $dW(t)$ is standard white noise. The OU process has an autocorrelation function
\begin{equation}
    \big\langle \delta_i(t)\,\delta_j(t') \big\rangle
    = \Lambda\, e^{-|t-t'|/\tau_c}\,\delta_{ij},
\end{equation}
showing that $\tau_c$ is the correlation time of the noise and $\Lambda$ is its variance.
From this form, the limits are immediate: $\tau_c \to 0$ yields temporally uncorrelated (white) noise, whereas
$\tau_c \to \infty$ produces static disorder with variance $\Lambda$, as shown in \cref{fig:Noise_effect}a. 

Delocalisation within a disordered aggregate can be quantified by the participation ratio (PR),
\begin{equation}
    \mathrm{PR}(\psi) = \bigg(\sum_{\alpha} |C_\alpha|^4\bigg)^{-1},
\end{equation}
where $C_\alpha$ is the amplitude of the site-basis configuration~$\alpha$. For example, for a two-particle spin donor state expressed in the site basis as $\ket{\psi}=\sum_{i>j} c_{ij} \ket{D_i}\ket{D_j}$, $\mathrm{PR}(\psi) = \left(\sum_{i> j} |c_{ij}|^4\right)^{-1}$.
The PR ranges from 1 for a localised state to maximum values for fully delocalised states of $\tbinom{N_\D}{m_\D}$ for spins and $\tbinom{N_\D+R_\D-1}{R_\D}$ for HOs. 

In the presence of noise, the system evolution becomes stochastic, meaning that the transfer rate can no longer be determined from~\cref{eqn:Fermi_rate}. Instead, the instantaneous rate fluctuates due to the random modulation of site energies. To obtain an effective rate, we simulate the dynamics under many noise realisations up to time $t = 1/\gamma_0$ and compute the time-averaged collective emission rate $\gamma_0\langle J^+ J^- \rangle$. In the absence of noise, this term yields the Dicke emission rate in \cref{eqn:Dicke_rate}, but with noise present, it captures the influence of the stochastic modulation.

Before turning to intra-aggregate couplings, we note that $\gamma_0$ suppresses the deleterious influence of disorder and noise, allowing CEEAT to persist. CEEAT can occur in an aggregate as long as the noise amplitude $\Lambda$ is small compared to the decay rate $\gamma_0$~\cite{Schuurmans1982,Shahbazyan2000},
\begin{equation}
\label{eqn:natural_condition}
\Lambda \ll \gamma_0.
\end{equation}
Heuristically, a large $\gamma_0$ (corresponding to strong donor-acceptor coupling $V_{\D\A}$) means that donor states are short-lived, making it impossible to distinguish their energies on scales smaller than $\gamma_0$. In the case of SR, this corresponds to the case of overlapping natural linewidths. Therefore, if site energies differ (statically or dynamically) by an amount $\Lambda$ smaller than $\gamma_0$, they can still support interfering pathways, resulting in CEEAT. 

The persistence of CEEAT due to $\gamma_0$ is seen in~\cref{fig:Noise_effect}b, where there is an enhancement even at $V_{\D\D}=0$, provided~\cref{eqn:natural_condition} is met. Indeed, partial CEEAT, under the general definition, can occur so long as $\Lambda \lesssim \gamma_0$.

Intra-aggregate coupling can be modelled by adding a term to the spin Hamiltonian of the form
\begin{equation}
    H_\mathrm{intra}^\mathrm{spin} = \sum_{i< j}^{N} V_{ij} (\sigma_i^+\sigma_j^-+\sigma_i^-\sigma_j^+ ),
\end{equation}
while, for an HO aggregate, it could take the form
\begin{equation}
    H_\mathrm{intra}^\mathrm{HO} = \sum_{i < j}^{N} V_{ij}(a_i^\dagger a_j+a_j^\dg a_i).
\end{equation}
To recover near-maximal CEEAT enhancement in the Dicke limit even in the presence of disorder, the system’s couplings must preserve permutational symmetry; e.g., they could be equal all-to-all couplings, equal couplings among sites arranged in a ring, or equal nearest-neighbour couplings in infinite lattices. In our calculations, we assume equal all-to-all couplings, denoted $V_{\D\D}$ within a donor aggregate or $V_{\A\A}$ within an acceptor. However, even disordered couplings have a partial protective effect, as we discuss below.

Intra-aggregate couplings increase the delocalisation of states, counteracting the localising effects of disorder~\cite{Celardo2014_static} and noise~\cite{Celardo2014_dynamic}. \Cref{fig:PR_noise} shows the effect of disorder and couplings on the average double-excitation PR in a spin aggregate of four sites. At small couplings, disorder localises the states close to $\mathrm{PR}=1$, an effect that is reversed by sufficiently strong coupling; for $V\gg  \Lambda$, the states have the maximum possible PR of 6.

Introducing intra-aggregate couplings is a generic way to improve the robustness of all CEEAT effects. In particular, increasing the intra-aggregate couplings allows larger amounts of disorder and noise before the CEEAT enhancement diminishes. In other words, the intra-aggregate couplings create another way (in addition to \cref{eqn:natural_condition}) to preserve CEEAT; in particular, maximum enhancement in the donor can be recovered in the limit
\begin{equation}
\label{eqn:Deloc_Rule}
V_{\D\D} \gg \Lambda_\D,\tau_\mathrm{c}^{-1},
\end{equation}
as shown in~\cref{fig:Noise_effect}b,c. 
An equivalent expression applies for intra-aggregate couplings in the acceptor. \Cref{eqn:Deloc_Rule} states that CEEAT is maintained if the noise is both weak and slow compared to the intra-aggregate couplings. If the condition $V_{\D\D} \gg \tau_{\mathrm{c}}^{-1}$ is not satisfied, the rapid modulation of the site energies rapidly scrambles the relative phases and suppresses the cooperative response, even when $V_{\D\D} \gg \Lambda_\D$.

A system can still exhibit non-maximal CEEAT even if the inequalities in \cref{eqn:Deloc_Rule} are only approximately satisfied (i.e., $\gg$ is replaced with $\gtrsim$). In this regime, the intra-aggregate couplings remain large enough to partially delocalise the excitations, maintaining interfering transfer pathways even though the enhancement is reduced from its maximum possible value.

The restoration of the maximum possible PR in \cref{fig:PR_noise} and the maximum possible rate in \cref{fig:Noise_effect}b,c is a consequence of our use of symmetric all-to-all intra-donor couplings. If those couplings were disordered (unequal), the resulting eigenstates would not be fully delocalised. Nevertheless, partial delocalisation and partial CEEAT (consistent with the general definition) remain in the case of disordered couplings. In the presence of disorder, any amount of coupling always increases delocalisation, therefore enabling at least partial CEEAT.

\section{Discussion}

We have presented a unified picture of CEEAT effects by highlighting their common origin in delocalisation-enhanced and delocalisation-suppressed quantum dynamics. This includes a description of new effects, such as HO SA or the mixed ST between spin and HO aggregates. The consistent Dicke treatment has given the scalings of all of these effects with system size, summarised in \cref{tab:Collective_effect}. These include unbounded rates for processes involving highly excited HO aggregates. Furthermore, we showed that all of these effects can be engineered for robustness against disorder and noise using intra-aggregate couplings.

We anticipate that other processes can be incorporated into our framework; two straightforward extensions could include processes that modulate CEEAT as well as hybrid processes involving multiple types of CEEAT.

By modulation, we mean the introduction of additional degrees of freedom that modify the parameters of the CEEAT Hamiltonian without being donors or acceptors themselves, such as phonons~\cite{Wiercinski2024} or nuclear motion~\cite{Tao2025}. For example, allowing donors and acceptors access to shared cavity or phonon modes can mediate interactions between them~\cite{Pallmann2024}, effectively increasing the donor-acceptor coupling. In molecular systems, a similar effect is superexchange-mediated bridged charge transfer~\cite{Petrov2002,Nitzan2006,Taylor2018}, where molecules between the donor and the acceptor supply virtual states that can facilitate donor-acceptor tunnelling by effectively increasing their coupling. Couplings and rates can also be affected by engineering the electromagnetic environment~\cite{Brown2019,Burgess2025}.

By hybrid processes, we mean situations where multiple CEEAT processes can occur within a single system, either sequentially or simultaneously. The analysis remains straightforward, because the enhancements apply to each rate separately. An example of hybrid CEEAT is collective energy migration~\cite{Dias2021}, a transfer process similar to ST, but arising not from direct coupling but from a combination of SR and SA.

There are also opportunities to discover new CEEAT phenomena if our approach is applied to collective processes that are not described by rate equations. For example, if donor-acceptor coupling is strong, it becomes impossible to neatly separate the donors from the acceptors. Examples where this situation occurs include polariton-assisted transfer (PARET)~\cite{Du2018,Ribeiro2018}, where a cavity facilitates strong coupling, and superradiance ringing~\cite{Skribanowitz1973}, the repeated collective emission and reabsorption of excitations. Some of these processes may transition to a rate process under sufficient damping; for example, sufficiently damped SR ringing~\cite{Kaluzny1983} can be regarded as a form of mixed ST from spins to HOs. 

We also anticipate the application of our framework to driven or steady-state systems, in distinction to the initial-state problem we addressed above. For instance, steady-state superradiant emission---such as in a superradiant laser~\cite{Bohnet2012}, where an ensemble of spins is continuously pumped and collectively emits into a cavity---could be generalised to ensembles of other degrees of freedom. In driven scenarios, collective behaviour manifests primarily in the frequency domain, where the interplay of driving, dissipation, and disorder can yield characteristic signatures including linewidth narrowing or broadening~\cite{Zhou2015}.

Finally, we look forward to the experimental demonstration of some of the proposed processes and high-order scaling. We expect that the robustness supplied by intra-aggregate couplings will make experimental demonstrations easier, particularly in large systems. Sufficient intra-aggregate couplings would relax the need for strict permutation symmetry, enabling robust collective behaviour under more general and experimentally accessible conditions. Because the ideas above apply to most quantum degrees of freedom, they could likely be realised in a wide range of platforms---including superconducting circuits, trapped ions, Rydberg atom arrays, and molecular aggregates---offering exciting prospects for building robust and scalable quantum technologies involving CEEAT effects.

\section*{Acknowledgements} 
A.K.~and I.K.~were supported by the Australian Research Council (FT230100653) and a Sydney Quantum Academy scholarship. E.M.G.~acknowledges support from The Leverhulme Trust (Grant No. RPG-2022-335) and Innovate UK (Grant No. 10120741).
    
\bibliography{bib}

\begin{thebibliography}{92}%
\makeatletter
\providecommand \@ifxundefined [1]{%
 \@ifx{#1\undefined}
}%
\providecommand \@ifnum [1]{%
 \ifnum #1\expandafter \@firstoftwo
 \else \expandafter \@secondoftwo
 \fi
}%
\providecommand \@ifx [1]{%
 \ifx #1\expandafter \@firstoftwo
 \else \expandafter \@secondoftwo
 \fi
}%
\providecommand \natexlab [1]{#1}%
\providecommand \enquote  [1]{``#1''}%
\providecommand \bibnamefont  [1]{#1}%
\providecommand \bibfnamefont [1]{#1}%
\providecommand \citenamefont [1]{#1}%
\providecommand \href@noop [0]{\@secondoftwo}%
\providecommand \href [0]{\begingroup \@sanitize@url \@href}%
\providecommand \@href[1]{\@@startlink{#1}\@@href}%
\providecommand \@@href[1]{\endgroup#1\@@endlink}%
\providecommand \@sanitize@url [0]{\catcode `\\12\catcode `\$12\catcode `\&12\catcode `\#12\catcode `\^12\catcode `\_12\catcode `\%12\relax}%
\providecommand \@@startlink[1]{}%
\providecommand \@@endlink[0]{}%
\providecommand \url  [0]{\begingroup\@sanitize@url \@url }%
\providecommand \@url [1]{\endgroup\@href {#1}{\urlprefix }}%
\providecommand \urlprefix  [0]{URL }%
\providecommand \Eprint [0]{\href }%
\providecommand \doibase [0]{https://doi.org/}%
\providecommand \selectlanguage [0]{\@gobble}%
\providecommand \bibinfo  [0]{\@secondoftwo}%
\providecommand \bibfield  [0]{\@secondoftwo}%
\providecommand \translation [1]{[#1]}%
\providecommand \BibitemOpen [0]{}%
\providecommand \bibitemStop [0]{}%
\providecommand \bibitemNoStop [0]{.\EOS\space}%
\providecommand \EOS [0]{\spacefactor3000\relax}%
\providecommand \BibitemShut  [1]{\csname bibitem#1\endcsname}%
\let\auto@bib@innerbib\@empty
\bibitem [{\citenamefont {Dicke}(1954)}]{Dicke1954}%
  \BibitemOpen
  \bibfield  {author} {\bibinfo {author} {\bibfnamefont {R.~H.}\ \bibnamefont {Dicke}},\ }\bibfield  {title} {\bibinfo {title} {Coherence in {S}pontaneous {R}adiation {P}rocesses},\ }\href {https://doi.org/10.1103/PhysRev.93.99} {\bibfield  {journal} {\bibinfo  {journal} {Phys. Rev.}\ }\textbf {\bibinfo {volume} {93}},\ \bibinfo {pages} {99} (\bibinfo {year} {1954})}\BibitemShut {NoStop}%
\bibitem [{\citenamefont {Higgins}\ \emph {et~al.}(2014)\citenamefont {Higgins}, \citenamefont {Benjamin}, \citenamefont {Stace}, \citenamefont {Milburn}, \citenamefont {Lovett},\ and\ \citenamefont {Gauger}}]{Higgins2014}%
  \BibitemOpen
  \bibfield  {author} {\bibinfo {author} {\bibfnamefont {K.~D.~B.}\ \bibnamefont {Higgins}}, \bibinfo {author} {\bibfnamefont {S.~C.}\ \bibnamefont {Benjamin}}, \bibinfo {author} {\bibfnamefont {T.~M.}\ \bibnamefont {Stace}}, \bibinfo {author} {\bibfnamefont {G.~J.}\ \bibnamefont {Milburn}}, \bibinfo {author} {\bibfnamefont {B.~W.}\ \bibnamefont {Lovett}},\ and\ \bibinfo {author} {\bibfnamefont {E.~M.}\ \bibnamefont {Gauger}},\ }\bibfield  {title} {\bibinfo {title} {Superabsorption of {L}ight via {Q}uantum {E}ngineering},\ }\href {https://doi.org/10.1038/ncomms5705} {\bibfield  {journal} {\bibinfo  {journal} {Nat. Commun.}\ }\textbf {\bibinfo {volume} {5}},\ \bibinfo {pages} {4705} (\bibinfo {year} {2014})}\BibitemShut {NoStop}%
\bibitem [{\citenamefont {Str\c{e}k}(1977)}]{Strk1977}%
  \BibitemOpen
  \bibfield  {author} {\bibinfo {author} {\bibfnamefont {W.}~\bibnamefont {Str\c{e}k}},\ }\bibfield  {title} {\bibinfo {title} {Cooperative energy transfer},\ }\href {https://doi.org/10.1016/0375-9601(77)90427-3} {\bibfield  {journal} {\bibinfo  {journal} {Phys. Lett. A}\ }\textbf {\bibinfo {volume} {62}},\ \bibinfo {pages} {315} (\bibinfo {year} {1977})}\BibitemShut {NoStop}%
\bibitem [{\citenamefont {Lloyd}\ and\ \citenamefont {Mohseni}(2010)}]{Lloyd2010}%
  \BibitemOpen
  \bibfield  {author} {\bibinfo {author} {\bibfnamefont {S.}~\bibnamefont {Lloyd}}\ and\ \bibinfo {author} {\bibfnamefont {M.}~\bibnamefont {Mohseni}},\ }\bibfield  {title} {\bibinfo {title} {Symmetry-{E}nhanced {S}upertransfer of {D}elocalized {Q}uantum {S}tates},\ }\href {https://doi.org/10.1088/1367-2630/12/7/075020} {\bibfield  {journal} {\bibinfo  {journal} {New J. Phys.}\ }\textbf {\bibinfo {volume} {12}},\ \bibinfo {pages} {075020} (\bibinfo {year} {2010})}\BibitemShut {NoStop}%
\bibitem [{\citenamefont {Liu}\ \emph {et~al.}(2023)\citenamefont {Liu}, \citenamefont {Shi}, \citenamefont {Yokoyama}, \citenamefont {Inoue}, \citenamefont {Sunaba}, \citenamefont {Oshikiri}, \citenamefont {Sun}, \citenamefont {Tamura}, \citenamefont {Ishihara}, \citenamefont {Sasaki},\ and\ \citenamefont {Misawa}}]{Liu2023}%
  \BibitemOpen
  \bibfield  {author} {\bibinfo {author} {\bibfnamefont {Y.-E.}\ \bibnamefont {Liu}}, \bibinfo {author} {\bibfnamefont {X.}~\bibnamefont {Shi}}, \bibinfo {author} {\bibfnamefont {T.}~\bibnamefont {Yokoyama}}, \bibinfo {author} {\bibfnamefont {S.}~\bibnamefont {Inoue}}, \bibinfo {author} {\bibfnamefont {Y.}~\bibnamefont {Sunaba}}, \bibinfo {author} {\bibfnamefont {T.}~\bibnamefont {Oshikiri}}, \bibinfo {author} {\bibfnamefont {Q.}~\bibnamefont {Sun}}, \bibinfo {author} {\bibfnamefont {M.}~\bibnamefont {Tamura}}, \bibinfo {author} {\bibfnamefont {H.}~\bibnamefont {Ishihara}}, \bibinfo {author} {\bibfnamefont {K.}~\bibnamefont {Sasaki}},\ and\ \bibinfo {author} {\bibfnamefont {H.}~\bibnamefont {Misawa}},\ }\bibfield  {title} {\bibinfo {title} {Quantum-{C}oherence-{E}nhanced {H}ot-{E}lectron {I}njection under {M}odal {S}trong {C}oupling},\ }\href {https://doi.org/10.1021/acsnano.2c12670} {\bibfield  {journal} {\bibinfo  {journal} {ACS Nano}\ }\textbf {\bibinfo {volume} {17}},\ \bibinfo {pages} {8315–8323}
  (\bibinfo {year} {2023})}\BibitemShut {NoStop}%
\bibitem [{\citenamefont {Meiser}\ \emph {et~al.}(2009)\citenamefont {Meiser}, \citenamefont {Ye}, \citenamefont {Carlson},\ and\ \citenamefont {Holland}}]{Meiser2009}%
  \BibitemOpen
  \bibfield  {author} {\bibinfo {author} {\bibfnamefont {D.}~\bibnamefont {Meiser}}, \bibinfo {author} {\bibfnamefont {J.}~\bibnamefont {Ye}}, \bibinfo {author} {\bibfnamefont {D.~R.}\ \bibnamefont {Carlson}},\ and\ \bibinfo {author} {\bibfnamefont {M.~J.}\ \bibnamefont {Holland}},\ }\bibfield  {title} {\bibinfo {title} {Prospects for a {M}illihertz-{L}inewidth {L}aser},\ }\href {https://doi.org/10.1103/PhysRevLett.102.163601} {\bibfield  {journal} {\bibinfo  {journal} {Phys. Rev. Lett.}\ }\textbf {\bibinfo {volume} {102}},\ \bibinfo {pages} {163601} (\bibinfo {year} {2009})}\BibitemShut {NoStop}%
\bibitem [{\citenamefont {Gorlach}\ \emph {et~al.}(2024)\citenamefont {Gorlach}, \citenamefont {Reinhardt}, \citenamefont {Pizzi}, \citenamefont {Ruimy}, \citenamefont {Baranes}, \citenamefont {Rivera},\ and\ \citenamefont {Kaminer}}]{Gorlach2024}%
  \BibitemOpen
  \bibfield  {author} {\bibinfo {author} {\bibfnamefont {A.}~\bibnamefont {Gorlach}}, \bibinfo {author} {\bibfnamefont {O.}~\bibnamefont {Reinhardt}}, \bibinfo {author} {\bibfnamefont {A.}~\bibnamefont {Pizzi}}, \bibinfo {author} {\bibfnamefont {R.}~\bibnamefont {Ruimy}}, \bibinfo {author} {\bibfnamefont {G.}~\bibnamefont {Baranes}}, \bibinfo {author} {\bibfnamefont {N.}~\bibnamefont {Rivera}},\ and\ \bibinfo {author} {\bibfnamefont {I.}~\bibnamefont {Kaminer}},\ }\bibfield  {title} {\bibinfo {title} {Double-superradiant cathodoluminescence},\ }\href {https://doi.org/10.1103/PhysRevA.109.023722} {\bibfield  {journal} {\bibinfo  {journal} {Phys. Rev. A}\ }\textbf {\bibinfo {volume} {109}},\ \bibinfo {pages} {023722} (\bibinfo {year} {2024})}\BibitemShut {NoStop}%
\bibitem [{\citenamefont {Zhu}\ \emph {et~al.}(2024{\natexlab{a}})\citenamefont {Zhu}, \citenamefont {Zhou}, \citenamefont {Wang}, \citenamefont {Thomas}, \citenamefont {Maity}, \citenamefont {Gutiérrez-Arzaluz}, \citenamefont {Wu}, \citenamefont {Sun}, \citenamefont {Jin}, \citenamefont {Cai}, \citenamefont {Wang}, \citenamefont {Alshareef}, \citenamefont {Bakr}, \citenamefont {Zhu},\ and\ \citenamefont {Mohammed}}]{Zhu2024}%
  \BibitemOpen
  \bibfield  {author} {\bibinfo {author} {\bibfnamefont {X.}~\bibnamefont {Zhu}}, \bibinfo {author} {\bibfnamefont {R.}~\bibnamefont {Zhou}}, \bibinfo {author} {\bibfnamefont {Z.}~\bibnamefont {Wang}}, \bibinfo {author} {\bibfnamefont {S.}~\bibnamefont {Thomas}}, \bibinfo {author} {\bibfnamefont {P.}~\bibnamefont {Maity}}, \bibinfo {author} {\bibfnamefont {L.}~\bibnamefont {Gutiérrez-Arzaluz}}, \bibinfo {author} {\bibfnamefont {W.}~\bibnamefont {Wu}}, \bibinfo {author} {\bibfnamefont {T.}~\bibnamefont {Sun}}, \bibinfo {author} {\bibfnamefont {T.}~\bibnamefont {Jin}}, \bibinfo {author} {\bibfnamefont {H.}~\bibnamefont {Cai}}, \bibinfo {author} {\bibfnamefont {J.-X.}\ \bibnamefont {Wang}}, \bibinfo {author} {\bibfnamefont {H.~N.}\ \bibnamefont {Alshareef}}, \bibinfo {author} {\bibfnamefont {O.~M.}\ \bibnamefont {Bakr}}, \bibinfo {author} {\bibfnamefont {Y.}~\bibnamefont {Zhu}},\ and\ \bibinfo {author} {\bibfnamefont {O.~F.}\ \bibnamefont {Mohammed}},\ }\bibfield  {title} {\bibinfo {title}
  {Lanthanide-{M}etal-{D}oped {L}ight-{H}arvesting {Q}uantum {D}ots for {E}xceptional {X}-ray {I}maging {S}cintillators},\ }\href {https://doi.org/10.1021/acsenergylett.4c02155} {\bibfield  {journal} {\bibinfo  {journal} {ACS Energy Lett.}\ }\textbf {\bibinfo {volume} {9}},\ \bibinfo {pages} {5137–5144} (\bibinfo {year} {2024}{\natexlab{a}})}\BibitemShut {NoStop}%
\bibitem [{\citenamefont {Andolina}\ \emph {et~al.}(2019)\citenamefont {Andolina}, \citenamefont {Keck}, \citenamefont {Mari}, \citenamefont {Giovannetti},\ and\ \citenamefont {Polini}}]{Andolina2019}%
  \BibitemOpen
  \bibfield  {author} {\bibinfo {author} {\bibfnamefont {G.~M.}\ \bibnamefont {Andolina}}, \bibinfo {author} {\bibfnamefont {M.}~\bibnamefont {Keck}}, \bibinfo {author} {\bibfnamefont {A.}~\bibnamefont {Mari}}, \bibinfo {author} {\bibfnamefont {V.}~\bibnamefont {Giovannetti}},\ and\ \bibinfo {author} {\bibfnamefont {M.}~\bibnamefont {Polini}},\ }\bibfield  {title} {\bibinfo {title} {Quantum versus classical many-body batteries},\ }\href {https://doi.org/10.1103/PhysRevB.99.205437} {\bibfield  {journal} {\bibinfo  {journal} {Phys. Rev. B}\ }\textbf {\bibinfo {volume} {99}},\ \bibinfo {pages} {205437} (\bibinfo {year} {2019})}\BibitemShut {NoStop}%
\bibitem [{\citenamefont {Quach}\ \emph {et~al.}(2022)\citenamefont {Quach}, \citenamefont {McGhee}, \citenamefont {Ganzer}, \citenamefont {Rouse}, \citenamefont {Lovett}, \citenamefont {Gauger}, \citenamefont {Keeling}, \citenamefont {Cerullo}, \citenamefont {Lidzey},\ and\ \citenamefont {Virgili}}]{Quach2022}%
  \BibitemOpen
  \bibfield  {author} {\bibinfo {author} {\bibfnamefont {J.~Q.}\ \bibnamefont {Quach}}, \bibinfo {author} {\bibfnamefont {K.~E.}\ \bibnamefont {McGhee}}, \bibinfo {author} {\bibfnamefont {L.}~\bibnamefont {Ganzer}}, \bibinfo {author} {\bibfnamefont {D.~M.}\ \bibnamefont {Rouse}}, \bibinfo {author} {\bibfnamefont {B.~W.}\ \bibnamefont {Lovett}}, \bibinfo {author} {\bibfnamefont {E.~M.}\ \bibnamefont {Gauger}}, \bibinfo {author} {\bibfnamefont {J.}~\bibnamefont {Keeling}}, \bibinfo {author} {\bibfnamefont {G.}~\bibnamefont {Cerullo}}, \bibinfo {author} {\bibfnamefont {D.~G.}\ \bibnamefont {Lidzey}},\ and\ \bibinfo {author} {\bibfnamefont {T.}~\bibnamefont {Virgili}},\ }\bibfield  {title} {\bibinfo {title} {Superabsorption in an organic microcavity: {T}oward a quantum battery},\ }\href {https://doi.org/10.1126/sciadv.abk3160} {\bibfield  {journal} {\bibinfo  {journal} {Sci. Adv.}\ }\textbf {\bibinfo {volume} {8}},\ \bibinfo {pages} {eabk3160} (\bibinfo {year} {2022})}\BibitemShut {NoStop}%
\bibitem [{\citenamefont {Campaioli}\ \emph {et~al.}(2024)\citenamefont {Campaioli}, \citenamefont {Gherardini}, \citenamefont {Quach}, \citenamefont {Polini},\ and\ \citenamefont {Andolina}}]{Campaioli2024}%
  \BibitemOpen
  \bibfield  {author} {\bibinfo {author} {\bibfnamefont {F.}~\bibnamefont {Campaioli}}, \bibinfo {author} {\bibfnamefont {S.}~\bibnamefont {Gherardini}}, \bibinfo {author} {\bibfnamefont {J.~Q.}\ \bibnamefont {Quach}}, \bibinfo {author} {\bibfnamefont {M.}~\bibnamefont {Polini}},\ and\ \bibinfo {author} {\bibfnamefont {G.~M.}\ \bibnamefont {Andolina}},\ }\bibfield  {title} {\bibinfo {title} {Colloquium: {Q}uantum batteries},\ }\href {https://doi.org/10.1103/RevModPhys.96.031001} {\bibfield  {journal} {\bibinfo  {journal} {Rev. Mod. Phys.}\ }\textbf {\bibinfo {volume} {96}},\ \bibinfo {pages} {031001} (\bibinfo {year} {2024})}\BibitemShut {NoStop}%
\bibitem [{\citenamefont {Dias}\ \emph {et~al.}(2024)\citenamefont {Dias}, \citenamefont {Wang}, \citenamefont {Nemoto}, \citenamefont {Nori},\ and\ \citenamefont {Munro}}]{Dias2024}%
  \BibitemOpen
  \bibfield  {author} {\bibinfo {author} {\bibfnamefont {J.}~\bibnamefont {Dias}}, \bibinfo {author} {\bibfnamefont {H.}~\bibnamefont {Wang}}, \bibinfo {author} {\bibfnamefont {K.}~\bibnamefont {Nemoto}}, \bibinfo {author} {\bibfnamefont {F.}~\bibnamefont {Nori}},\ and\ \bibinfo {author} {\bibfnamefont {W.~J.}\ \bibnamefont {Munro}},\ }\href {https://arxiv.org/abs/2410.19303} {\bibinfo {title} {Efficient charging of multiple open quantum batteries through dissipation and pumping}} (\bibinfo {year} {2024}),\ \Eprint {https://arxiv.org/abs/2410.19303} {arXiv:2410.19303 [quant-ph]} \BibitemShut {NoStop}%
\bibitem [{\citenamefont {Lvovsky}\ \emph {et~al.}(2009)\citenamefont {Lvovsky}, \citenamefont {Sanders},\ and\ \citenamefont {Tittel}}]{Lvovsky2009}%
  \BibitemOpen
  \bibfield  {author} {\bibinfo {author} {\bibfnamefont {A.~I.}\ \bibnamefont {Lvovsky}}, \bibinfo {author} {\bibfnamefont {B.~C.}\ \bibnamefont {Sanders}},\ and\ \bibinfo {author} {\bibfnamefont {W.}~\bibnamefont {Tittel}},\ }\bibfield  {title} {\bibinfo {title} {Optical quantum memory},\ }\href {https://doi.org/10.1038/nphoton.2009.231} {\bibfield  {journal} {\bibinfo  {journal} {Nat. Photonics}\ }\textbf {\bibinfo {volume} {3}},\ \bibinfo {pages} {706–714} (\bibinfo {year} {2009})}\BibitemShut {NoStop}%
\bibitem [{\citenamefont {McGuyer}\ \emph {et~al.}(2014)\citenamefont {McGuyer}, \citenamefont {McDonald}, \citenamefont {Iwata}, \citenamefont {Tarallo}, \citenamefont {Skomorowski}, \citenamefont {Moszynski},\ and\ \citenamefont {Zelevinsky}}]{McGuyer2014}%
  \BibitemOpen
  \bibfield  {author} {\bibinfo {author} {\bibfnamefont {B.~H.}\ \bibnamefont {McGuyer}}, \bibinfo {author} {\bibfnamefont {M.}~\bibnamefont {McDonald}}, \bibinfo {author} {\bibfnamefont {G.~Z.}\ \bibnamefont {Iwata}}, \bibinfo {author} {\bibfnamefont {M.~G.}\ \bibnamefont {Tarallo}}, \bibinfo {author} {\bibfnamefont {W.}~\bibnamefont {Skomorowski}}, \bibinfo {author} {\bibfnamefont {R.}~\bibnamefont {Moszynski}},\ and\ \bibinfo {author} {\bibfnamefont {T.}~\bibnamefont {Zelevinsky}},\ }\bibfield  {title} {\bibinfo {title} {Precise study of asymptotic physics with subradiant ultracold molecules},\ }\href {https://doi.org/10.1038/nphys3182} {\bibfield  {journal} {\bibinfo  {journal} {Nat. Phys.}\ }\textbf {\bibinfo {volume} {11}},\ \bibinfo {pages} {32–36} (\bibinfo {year} {2014})}\BibitemShut {NoStop}%
\bibitem [{\citenamefont {Pasquiou}(2014)}]{Pasquiou2014}%
  \BibitemOpen
  \bibfield  {author} {\bibinfo {author} {\bibfnamefont {B.}~\bibnamefont {Pasquiou}},\ }\bibfield  {title} {\bibinfo {title} {Subradiance spectroscopy},\ }\href {https://doi.org/10.1038/nphys3208} {\bibfield  {journal} {\bibinfo  {journal} {Nat. Phys.}\ }\textbf {\bibinfo {volume} {11}},\ \bibinfo {pages} {14–15} (\bibinfo {year} {2014})}\BibitemShut {NoStop}%
\bibitem [{\citenamefont {McDonald}(2017)}]{McDonald2017}%
  \BibitemOpen
  \bibfield  {author} {\bibinfo {author} {\bibfnamefont {M.}~\bibnamefont {McDonald}},\ }\bibinfo {title} {Subradiant {S}pectroscopy},\ in\ \href {https://doi.org/10.1007/978-3-319-68735-3_6} {\emph {\bibinfo {booktitle} {High {P}recision {O}ptical {S}pectroscopy and {Q}uantum {S}tate {S}elected {P}hotodissociation of {U}ltracold 88Sr2 {M}olecules in an {O}ptical {L}attice}}}\ (\bibinfo  {publisher} {Springer International Publishing},\ \bibinfo {year} {2017})\ p.\ \bibinfo {pages} {107–134}\BibitemShut {NoStop}%
\bibitem [{\citenamefont {Baghbanzadeh}\ and\ \citenamefont {Kassal}(2016)}]{Baghbanzadeh2016}%
  \BibitemOpen
  \bibfield  {author} {\bibinfo {author} {\bibfnamefont {S.}~\bibnamefont {Baghbanzadeh}}\ and\ \bibinfo {author} {\bibfnamefont {I.}~\bibnamefont {Kassal}},\ }\bibfield  {title} {\bibinfo {title} {Geometry, {S}upertransfer, and {O}ptimality in the {L}ight {H}arvesting of {P}urple {B}acteria},\ }\href {https://doi.org/10.1021/acs.jpclett.6b01779} {\bibfield  {journal} {\bibinfo  {journal} {J. Phys. Chem. Lett.}\ }\textbf {\bibinfo {volume} {7}},\ \bibinfo {pages} {3804–3811} (\bibinfo {year} {2016})}\BibitemShut {NoStop}%
\bibitem [{\citenamefont {Hepp}\ and\ \citenamefont {Lieb}(1973)}]{Hepp1973_a}%
  \BibitemOpen
  \bibfield  {author} {\bibinfo {author} {\bibfnamefont {K.}~\bibnamefont {Hepp}}\ and\ \bibinfo {author} {\bibfnamefont {E.~H.}\ \bibnamefont {Lieb}},\ }\bibfield  {title} {\bibinfo {title} {On the {S}uperradiant {P}hase {T}ransition for {M}olecules in a {Q}uantized {R}adiation {F}ield: the {D}icke {M}aser {M}odel},\ }\href {https://doi.org/10.1016/0003-4916(73)90039-0} {\bibfield  {journal} {\bibinfo  {journal} {Ann. Phys.}\ }\textbf {\bibinfo {volume} {76}},\ \bibinfo {pages} {360–404} (\bibinfo {year} {1973})}\BibitemShut {NoStop}%
\bibitem [{\citenamefont {Wang}\ and\ \citenamefont {Hioe}(1973)}]{Wang1973}%
  \BibitemOpen
  \bibfield  {author} {\bibinfo {author} {\bibfnamefont {Y.~K.}\ \bibnamefont {Wang}}\ and\ \bibinfo {author} {\bibfnamefont {F.~T.}\ \bibnamefont {Hioe}},\ }\bibfield  {title} {\bibinfo {title} {Phase {T}ransition in the {D}icke {M}odel of {S}uperradiance},\ }\href {https://doi.org/10.1103/PhysRevA.7.831} {\bibfield  {journal} {\bibinfo  {journal} {Phys. Rev. A}\ }\textbf {\bibinfo {volume} {7}},\ \bibinfo {pages} {831} (\bibinfo {year} {1973})}\BibitemShut {NoStop}%
\bibitem [{\citenamefont {Kirton}\ \emph {et~al.}(2019)\citenamefont {Kirton}, \citenamefont {Roses}, \citenamefont {Keeling},\ and\ \citenamefont {Dalla~Torre}}]{Kirton2019}%
  \BibitemOpen
  \bibfield  {author} {\bibinfo {author} {\bibfnamefont {P.}~\bibnamefont {Kirton}}, \bibinfo {author} {\bibfnamefont {M.~M.}\ \bibnamefont {Roses}}, \bibinfo {author} {\bibfnamefont {J.}~\bibnamefont {Keeling}},\ and\ \bibinfo {author} {\bibfnamefont {E.~G.}\ \bibnamefont {Dalla~Torre}},\ }\bibfield  {title} {\bibinfo {title} {{Introduction to the Dicke Model: From Equilibrium to Nonequilibrium, and Vice Versa}},\ }\href {https://doi.org/https://doi.org/10.1002/qute.201800043} {\bibfield  {journal} {\bibinfo  {journal} {Adv. Quantum Technol.}\ }\textbf {\bibinfo {volume} {2}},\ \bibinfo {pages} {1800043} (\bibinfo {year} {2019})}\BibitemShut {NoStop}%
\bibitem [{\citenamefont {Freedhoff}\ and\ \citenamefont {Kranendonk}(1967)}]{Freedhoff1967}%
  \BibitemOpen
  \bibfield  {author} {\bibinfo {author} {\bibfnamefont {H.}~\bibnamefont {Freedhoff}}\ and\ \bibinfo {author} {\bibfnamefont {J.~V.}\ \bibnamefont {Kranendonk}},\ }\bibfield  {title} {\bibinfo {title} {Theory of coherent resonant absorption and emission at infrared and optical frequencies},\ }\href {https://doi.org/10.1139/p67-142} {\bibfield  {journal} {\bibinfo  {journal} {Can. J. Phys.}\ }\textbf {\bibinfo {volume} {45}},\ \bibinfo {pages} {1833–1859} (\bibinfo {year} {1967})}\BibitemShut {NoStop}%
\bibitem [{\citenamefont {Gold}\ \emph {et~al.}(2025)\citenamefont {Gold}, \citenamefont {Saglam}, \citenamefont {Carpenter}, \citenamefont {Yadav}, \citenamefont {Beede}, \citenamefont {Walker}, \citenamefont {Saffman},\ and\ \citenamefont {Yavuz}}]{Gold2025}%
  \BibitemOpen
  \bibfield  {author} {\bibinfo {author} {\bibfnamefont {D.~C.}\ \bibnamefont {Gold}}, \bibinfo {author} {\bibfnamefont {U.}~\bibnamefont {Saglam}}, \bibinfo {author} {\bibfnamefont {S.}~\bibnamefont {Carpenter}}, \bibinfo {author} {\bibfnamefont {A.}~\bibnamefont {Yadav}}, \bibinfo {author} {\bibfnamefont {M.}~\bibnamefont {Beede}}, \bibinfo {author} {\bibfnamefont {T.~G.}\ \bibnamefont {Walker}}, \bibinfo {author} {\bibfnamefont {M.}~\bibnamefont {Saffman}},\ and\ \bibinfo {author} {\bibfnamefont {D.~D.}\ \bibnamefont {Yavuz}},\ }\bibfield  {title} {\bibinfo {title} {Experimental observation of subabsorption},\ }\href {https://doi.org/10.1103/6xnl-814s} {\bibfield  {journal} {\bibinfo  {journal} {Phys. Rev. A}\ }\textbf {\bibinfo {volume} {112}},\ \bibinfo {pages} {023708} (\bibinfo {year} {2025})}\BibitemShut {NoStop}%
\bibitem [{\citenamefont {Lidar}\ \emph {et~al.}(1998)\citenamefont {Lidar}, \citenamefont {Chuang},\ and\ \citenamefont {Whaley}}]{Lidar1998}%
  \BibitemOpen
  \bibfield  {author} {\bibinfo {author} {\bibfnamefont {D.~A.}\ \bibnamefont {Lidar}}, \bibinfo {author} {\bibfnamefont {I.~L.}\ \bibnamefont {Chuang}},\ and\ \bibinfo {author} {\bibfnamefont {K.~B.}\ \bibnamefont {Whaley}},\ }\bibfield  {title} {\bibinfo {title} {Decoherence-{F}ree {S}ubspaces for {Q}uantum {C}omputation},\ }\href {https://doi.org/10.1103/PhysRevLett.81.2594} {\bibfield  {journal} {\bibinfo  {journal} {Phys. Rev. Lett.}\ }\textbf {\bibinfo {volume} {81}},\ \bibinfo {pages} {2594} (\bibinfo {year} {1998})}\BibitemShut {NoStop}%
\bibitem [{\citenamefont {Kwiat}\ \emph {et~al.}(2000)\citenamefont {Kwiat}, \citenamefont {Berglund}, \citenamefont {Altepeter},\ and\ \citenamefont {White}}]{Kwiat2000}%
  \BibitemOpen
  \bibfield  {author} {\bibinfo {author} {\bibfnamefont {P.~G.}\ \bibnamefont {Kwiat}}, \bibinfo {author} {\bibfnamefont {A.~J.}\ \bibnamefont {Berglund}}, \bibinfo {author} {\bibfnamefont {J.~B.}\ \bibnamefont {Altepeter}},\ and\ \bibinfo {author} {\bibfnamefont {A.~G.}\ \bibnamefont {White}},\ }\bibfield  {title} {\bibinfo {title} {Experimental {V}erification of {D}ecoherence-{F}ree {S}ubspaces},\ }\href {https://doi.org/10.1126/science.290.5491.498} {\bibfield  {journal} {\bibinfo  {journal} {Science}\ }\textbf {\bibinfo {volume} {290}},\ \bibinfo {pages} {498–501} (\bibinfo {year} {2000})}\BibitemShut {NoStop}%
\bibitem [{\citenamefont {Sumi}(1998)}]{Sumi1998}%
  \BibitemOpen
  \bibfield  {author} {\bibinfo {author} {\bibfnamefont {H.}~\bibnamefont {Sumi}},\ }\bibfield  {title} {\bibinfo {title} {Theory on {R}ates of {E}xcitation-{E}nergy {T}ransfer between {M}olecular {A}ggregates through {D}istributed {T}ransition {D}ipoles with {A}pplication to the {A}ntenna {S}ystem in {B}acterial {P}hotosynthesis},\ }\href {https://doi.org/10.1021/jp983477u} {\bibfield  {journal} {\bibinfo  {journal} {J. Phys. Chem. B}\ }\textbf {\bibinfo {volume} {103}},\ \bibinfo {pages} {252–260} (\bibinfo {year} {1998})}\BibitemShut {NoStop}%
\bibitem [{\citenamefont {Taylor}\ and\ \citenamefont {Kassal}(2018)}]{Taylor2018}%
  \BibitemOpen
  \bibfield  {author} {\bibinfo {author} {\bibfnamefont {N.~B.}\ \bibnamefont {Taylor}}\ and\ \bibinfo {author} {\bibfnamefont {I.}~\bibnamefont {Kassal}},\ }\bibfield  {title} {\bibinfo {title} {Generalised {M}arcus theory for multi-molecular delocalised charge transfer},\ }\href {https://doi.org/10.1039/c8sc00053k} {\bibfield  {journal} {\bibinfo  {journal} {Chem. Sci.}\ }\textbf {\bibinfo {volume} {9}},\ \bibinfo {pages} {2942–2951} (\bibinfo {year} {2018})}\BibitemShut {NoStop}%
\bibitem [{\citenamefont {Skribanowitz}\ \emph {et~al.}(1973)\citenamefont {Skribanowitz}, \citenamefont {Herman}, \citenamefont {MacGillivray},\ and\ \citenamefont {Feld}}]{Skribanowitz1973}%
  \BibitemOpen
  \bibfield  {author} {\bibinfo {author} {\bibfnamefont {N.}~\bibnamefont {Skribanowitz}}, \bibinfo {author} {\bibfnamefont {I.~P.}\ \bibnamefont {Herman}}, \bibinfo {author} {\bibfnamefont {J.~C.}\ \bibnamefont {MacGillivray}},\ and\ \bibinfo {author} {\bibfnamefont {M.~S.}\ \bibnamefont {Feld}},\ }\bibfield  {title} {\bibinfo {title} {Observation of {D}icke {S}uperradiance in {O}ptically {P}umped {H}{F} {G}as},\ }\href {https://doi.org/10.1103/PhysRevLett.30.309} {\bibfield  {journal} {\bibinfo  {journal} {Phys. Rev. Lett.}\ }\textbf {\bibinfo {volume} {30}},\ \bibinfo {pages} {309} (\bibinfo {year} {1973})}\BibitemShut {NoStop}%
\bibitem [{\citenamefont {Yang}\ \emph {et~al.}(2021)\citenamefont {Yang}, \citenamefont {Oh}, \citenamefont {Han}, \citenamefont {Son}, \citenamefont {Kim}, \citenamefont {Kim}, \citenamefont {Lee},\ and\ \citenamefont {An}}]{Yang2021}%
  \BibitemOpen
  \bibfield  {author} {\bibinfo {author} {\bibfnamefont {D.}~\bibnamefont {Yang}}, \bibinfo {author} {\bibfnamefont {S.-h.}\ \bibnamefont {Oh}}, \bibinfo {author} {\bibfnamefont {J.}~\bibnamefont {Han}}, \bibinfo {author} {\bibfnamefont {G.}~\bibnamefont {Son}}, \bibinfo {author} {\bibfnamefont {J.}~\bibnamefont {Kim}}, \bibinfo {author} {\bibfnamefont {J.}~\bibnamefont {Kim}}, \bibinfo {author} {\bibfnamefont {M.}~\bibnamefont {Lee}},\ and\ \bibinfo {author} {\bibfnamefont {K.}~\bibnamefont {An}},\ }\bibfield  {title} {\bibinfo {title} {Realization of superabsorption by time reversal of superradiance},\ }\href {https://doi.org/10.1038/s41566-021-00770-6} {\bibfield  {journal} {\bibinfo  {journal} {Nat. Photonics}\ }\textbf {\bibinfo {volume} {15}},\ \bibinfo {pages} {272–276} (\bibinfo {year} {2021})}\BibitemShut {NoStop}%
\bibitem [{\citenamefont {Pavolini}\ \emph {et~al.}(1985)\citenamefont {Pavolini}, \citenamefont {Crubellier}, \citenamefont {Pillet}, \citenamefont {Cabaret},\ and\ \citenamefont {Liberman}}]{Pavolini1985}%
  \BibitemOpen
  \bibfield  {author} {\bibinfo {author} {\bibfnamefont {D.}~\bibnamefont {Pavolini}}, \bibinfo {author} {\bibfnamefont {A.}~\bibnamefont {Crubellier}}, \bibinfo {author} {\bibfnamefont {P.}~\bibnamefont {Pillet}}, \bibinfo {author} {\bibfnamefont {L.}~\bibnamefont {Cabaret}},\ and\ \bibinfo {author} {\bibfnamefont {S.}~\bibnamefont {Liberman}},\ }\bibfield  {title} {\bibinfo {title} {Experimental {E}vidence for {S}ubradiance},\ }\href {https://doi.org/10.1103/PhysRevLett.54.1917} {\bibfield  {journal} {\bibinfo  {journal} {Phys. Rev. Lett.}\ }\textbf {\bibinfo {volume} {54}},\ \bibinfo {pages} {1917} (\bibinfo {year} {1985})}\BibitemShut {NoStop}%
\bibitem [{\citenamefont {Guerin}\ \emph {et~al.}(2016)\citenamefont {Guerin}, \citenamefont {Ara\'ujo},\ and\ \citenamefont {Kaiser}}]{Guerin2016}%
  \BibitemOpen
  \bibfield  {author} {\bibinfo {author} {\bibfnamefont {W.}~\bibnamefont {Guerin}}, \bibinfo {author} {\bibfnamefont {M.~O.}\ \bibnamefont {Ara\'ujo}},\ and\ \bibinfo {author} {\bibfnamefont {R.}~\bibnamefont {Kaiser}},\ }\bibfield  {title} {\bibinfo {title} {Subradiance in a {L}arge {C}loud of {C}old {A}toms},\ }\href {https://doi.org/10.1103/PhysRevLett.116.083601} {\bibfield  {journal} {\bibinfo  {journal} {Phys. Rev. Lett.}\ }\textbf {\bibinfo {volume} {116}},\ \bibinfo {pages} {083601} (\bibinfo {year} {2016})}\BibitemShut {NoStop}%
\bibitem [{\citenamefont {Potočnik}\ \emph {et~al.}(2018)\citenamefont {Potočnik}, \citenamefont {Bargerbos}, \citenamefont {Schr\"{o}der}, \citenamefont {Khan}, \citenamefont {Collodo}, \citenamefont {Gasparinetti}, \citenamefont {Salathé}, \citenamefont {Creatore}, \citenamefont {Eichler}, \citenamefont {T\"{u}reci}, \citenamefont {Chin},\ and\ \citenamefont {Wallraff}}]{Potonik2018}%
  \BibitemOpen
  \bibfield  {author} {\bibinfo {author} {\bibfnamefont {A.}~\bibnamefont {Potočnik}}, \bibinfo {author} {\bibfnamefont {A.}~\bibnamefont {Bargerbos}}, \bibinfo {author} {\bibfnamefont {F.~A. Y.~N.}\ \bibnamefont {Schr\"{o}der}}, \bibinfo {author} {\bibfnamefont {S.~A.}\ \bibnamefont {Khan}}, \bibinfo {author} {\bibfnamefont {M.~C.}\ \bibnamefont {Collodo}}, \bibinfo {author} {\bibfnamefont {S.}~\bibnamefont {Gasparinetti}}, \bibinfo {author} {\bibfnamefont {Y.}~\bibnamefont {Salathé}}, \bibinfo {author} {\bibfnamefont {C.}~\bibnamefont {Creatore}}, \bibinfo {author} {\bibfnamefont {C.}~\bibnamefont {Eichler}}, \bibinfo {author} {\bibfnamefont {H.~E.}\ \bibnamefont {T\"{u}reci}}, \bibinfo {author} {\bibfnamefont {A.~W.}\ \bibnamefont {Chin}},\ and\ \bibinfo {author} {\bibfnamefont {A.}~\bibnamefont {Wallraff}},\ }\bibfield  {title} {\bibinfo {title} {Studying light-harvesting models with superconducting circuits},\ }\href {https://doi.org/10.1038/s41467-018-03312-x} {\bibfield  {journal} {\bibinfo
  {journal} {Nat. Commun.}\ }\textbf {\bibinfo {volume} {9}},\ \bibinfo {pages} {1} (\bibinfo {year} {2018})}\BibitemShut {NoStop}%
\bibitem [{\citenamefont {Kushwaha}\ and\ \citenamefont {Kassal}(2025)}]{Kushwaha2025}%
  \BibitemOpen
  \bibfield  {author} {\bibinfo {author} {\bibfnamefont {A.}~\bibnamefont {Kushwaha}}\ and\ \bibinfo {author} {\bibfnamefont {I.}~\bibnamefont {Kassal}},\ }\href {https://arxiv.org/abs/2506.05045} {\bibinfo {title} {Engineering {Q}uantum-{E}nhanced {T}ransport by {S}upertransfer}} (\bibinfo {year} {2025}),\ \Eprint {https://arxiv.org/abs/2506.05045} {arXiv:2506.05045 [quant-ph]} \BibitemShut {NoStop}%
\bibitem [{\citenamefont {Gross}\ and\ \citenamefont {Haroche}(1982)}]{Gross1982}%
  \BibitemOpen
  \bibfield  {author} {\bibinfo {author} {\bibfnamefont {M.}~\bibnamefont {Gross}}\ and\ \bibinfo {author} {\bibfnamefont {S.}~\bibnamefont {Haroche}},\ }\bibfield  {title} {\bibinfo {title} {Superradiance: {A}n essay on the theory of collective spontaneous emission},\ }\href {https://doi.org/10.1016/0370-1573(82)90102-8} {\bibfield  {journal} {\bibinfo  {journal} {Phys. Rep.}\ }\textbf {\bibinfo {volume} {93}},\ \bibinfo {pages} {301–396} (\bibinfo {year} {1982})}\BibitemShut {NoStop}%
\bibitem [{\citenamefont {Cong}\ \emph {et~al.}(2016)\citenamefont {Cong}, \citenamefont {Zhang}, \citenamefont {Wang}, \citenamefont {Noe}, \citenamefont {Belyanin},\ and\ \citenamefont {Kono}}]{Cong2016}%
  \BibitemOpen
  \bibfield  {author} {\bibinfo {author} {\bibfnamefont {K.}~\bibnamefont {Cong}}, \bibinfo {author} {\bibfnamefont {Q.}~\bibnamefont {Zhang}}, \bibinfo {author} {\bibfnamefont {Y.}~\bibnamefont {Wang}}, \bibinfo {author} {\bibfnamefont {G.~T.}\ \bibnamefont {Noe}}, \bibinfo {author} {\bibfnamefont {A.}~\bibnamefont {Belyanin}},\ and\ \bibinfo {author} {\bibfnamefont {J.}~\bibnamefont {Kono}},\ }\bibfield  {title} {\bibinfo {title} {Dicke superradiance in solids [{I}nvited]},\ }\href {https://doi.org/10.1364/josab.33.000c80} {\bibfield  {journal} {\bibinfo  {journal} {J. Opt. Soc. Am. B}\ }\textbf {\bibinfo {volume} {33}},\ \bibinfo {pages} {C80} (\bibinfo {year} {2016})}\BibitemShut {NoStop}%
\bibitem [{\citenamefont {Nefedkin}\ \emph {et~al.}(2017)\citenamefont {Nefedkin}, \citenamefont {Andrianov}, \citenamefont {Zyablovsky}, \citenamefont {Pukhov}, \citenamefont {Vinogradov},\ and\ \citenamefont {Lisyansky}}]{Nefedkin2017}%
  \BibitemOpen
  \bibfield  {author} {\bibinfo {author} {\bibfnamefont {N.~E.}\ \bibnamefont {Nefedkin}}, \bibinfo {author} {\bibfnamefont {E.~S.}\ \bibnamefont {Andrianov}}, \bibinfo {author} {\bibfnamefont {A.~A.}\ \bibnamefont {Zyablovsky}}, \bibinfo {author} {\bibfnamefont {A.~A.}\ \bibnamefont {Pukhov}}, \bibinfo {author} {\bibfnamefont {A.~P.}\ \bibnamefont {Vinogradov}},\ and\ \bibinfo {author} {\bibfnamefont {A.~A.}\ \bibnamefont {Lisyansky}},\ }\bibfield  {title} {\bibinfo {title} {Superradiance of non-{D}icke states},\ }\href {https://doi.org/10.1364/oe.25.002790} {\bibfield  {journal} {\bibinfo  {journal} {Opt. Express}\ }\textbf {\bibinfo {volume} {25}},\ \bibinfo {pages} {2790} (\bibinfo {year} {2017})}\BibitemShut {NoStop}%
\bibitem [{\citenamefont {Bojer}\ and\ \citenamefont {von Zanthier}(2022)}]{Bojer2022}%
  \BibitemOpen
  \bibfield  {author} {\bibinfo {author} {\bibfnamefont {M.}~\bibnamefont {Bojer}}\ and\ \bibinfo {author} {\bibfnamefont {J.}~\bibnamefont {von Zanthier}},\ }\bibfield  {title} {\bibinfo {title} {Dicke-like superradiance of distant noninteracting atoms},\ }\href {https://doi.org/10.1103/PhysRevA.106.053712} {\bibfield  {journal} {\bibinfo  {journal} {Phys. Rev. A}\ }\textbf {\bibinfo {volume} {106}},\ \bibinfo {pages} {053712} (\bibinfo {year} {2022})}\BibitemShut {NoStop}%
\bibitem [{\citenamefont {Sierra}\ \emph {et~al.}(2022)\citenamefont {Sierra}, \citenamefont {Masson},\ and\ \citenamefont {Asenjo-Garcia}}]{Sierra2022}%
  \BibitemOpen
  \bibfield  {author} {\bibinfo {author} {\bibfnamefont {E.}~\bibnamefont {Sierra}}, \bibinfo {author} {\bibfnamefont {S.~J.}\ \bibnamefont {Masson}},\ and\ \bibinfo {author} {\bibfnamefont {A.}~\bibnamefont {Asenjo-Garcia}},\ }\bibfield  {title} {\bibinfo {title} {Dicke {S}uperradiance in {O}rdered {L}attices: {D}imensionality {M}atters},\ }\href {https://doi.org/10.1103/physrevresearch.4.023207} {\bibfield  {journal} {\bibinfo  {journal} {Phys. Rev. Res.}\ }\textbf {\bibinfo {volume} {4}},\ \bibinfo {pages} {023207} (\bibinfo {year} {2022})}\BibitemShut {NoStop}%
\bibitem [{\citenamefont {Liedl}\ \emph {et~al.}(2024)\citenamefont {Liedl}, \citenamefont {Tebbenjohanns}, \citenamefont {Bach}, \citenamefont {Pucher}, \citenamefont {Rauschenbeutel},\ and\ \citenamefont {Schneeweiss}}]{Liedl2024}%
  \BibitemOpen
  \bibfield  {author} {\bibinfo {author} {\bibfnamefont {C.}~\bibnamefont {Liedl}}, \bibinfo {author} {\bibfnamefont {F.}~\bibnamefont {Tebbenjohanns}}, \bibinfo {author} {\bibfnamefont {C.}~\bibnamefont {Bach}}, \bibinfo {author} {\bibfnamefont {S.}~\bibnamefont {Pucher}}, \bibinfo {author} {\bibfnamefont {A.}~\bibnamefont {Rauschenbeutel}},\ and\ \bibinfo {author} {\bibfnamefont {P.}~\bibnamefont {Schneeweiss}},\ }\bibfield  {title} {\bibinfo {title} {Observation of {S}uperradiant {B}ursts in a {C}ascaded {Q}uantum {S}ystem},\ }\href {https://doi.org/10.1103/PhysRevX.14.011020} {\bibfield  {journal} {\bibinfo  {journal} {Phys. Rev. X}\ }\textbf {\bibinfo {volume} {14}},\ \bibinfo {pages} {011020} (\bibinfo {year} {2024})}\BibitemShut {NoStop}%
\bibitem [{\citenamefont {Holzinger}\ and\ \citenamefont {Yelin}(2025)}]{Holzinger2025}%
  \BibitemOpen
  \bibfield  {author} {\bibinfo {author} {\bibfnamefont {R.}~\bibnamefont {Holzinger}}\ and\ \bibinfo {author} {\bibfnamefont {S.~F.}\ \bibnamefont {Yelin}},\ }\href {https://doi.org/10.48550/ARXIV.2506.12649} {\bibinfo {title} {Beyond {D}icke superradiance: {U}niversal scaling of the peak emission rate}} (\bibinfo {year} {2025})\BibitemShut {NoStop}%
\bibitem [{\citenamefont {Scully}\ and\ \citenamefont {Svidzinsky}(2009)}]{Scully2009}%
  \BibitemOpen
  \bibfield  {author} {\bibinfo {author} {\bibfnamefont {M.~O.}\ \bibnamefont {Scully}}\ and\ \bibinfo {author} {\bibfnamefont {A.~A.}\ \bibnamefont {Svidzinsky}},\ }\bibfield  {title} {\bibinfo {title} {The {S}uper of {S}uperradiance},\ }\href {https://doi.org/10.1126/science.1176695} {\bibfield  {journal} {\bibinfo  {journal} {Science}\ }\textbf {\bibinfo {volume} {325}},\ \bibinfo {pages} {1510–1511} (\bibinfo {year} {2009})}\BibitemShut {NoStop}%
\bibitem [{\citenamefont {Spano}\ \emph {et~al.}(1991)\citenamefont {Spano}, \citenamefont {Kuklinski},\ and\ \citenamefont {Mukamel}}]{Spano1991}%
  \BibitemOpen
  \bibfield  {author} {\bibinfo {author} {\bibfnamefont {F.~C.}\ \bibnamefont {Spano}}, \bibinfo {author} {\bibfnamefont {J.~R.}\ \bibnamefont {Kuklinski}},\ and\ \bibinfo {author} {\bibfnamefont {S.}~\bibnamefont {Mukamel}},\ }\bibfield  {title} {\bibinfo {title} {Cooperative radiative dynamics in molecular aggregates},\ }\href {https://doi.org/10.1063/1.460185} {\bibfield  {journal} {\bibinfo  {journal} {J. Chem. Phys.}\ }\textbf {\bibinfo {volume} {94}},\ \bibinfo {pages} {7534–7544} (\bibinfo {year} {1991})}\BibitemShut {NoStop}%
\bibitem [{\citenamefont {Choudhary}\ \emph {et~al.}(2019)\citenamefont {Choudhary}, \citenamefont {De~Leon}, \citenamefont {Swiecicki}, \citenamefont {Awan}, \citenamefont {Schulz}, \citenamefont {Upham}, \citenamefont {Alam}, \citenamefont {Sipe},\ and\ \citenamefont {Boyd}}]{Choudhary2019}%
  \BibitemOpen
  \bibfield  {author} {\bibinfo {author} {\bibfnamefont {S.}~\bibnamefont {Choudhary}}, \bibinfo {author} {\bibfnamefont {I.}~\bibnamefont {De~Leon}}, \bibinfo {author} {\bibfnamefont {S.}~\bibnamefont {Swiecicki}}, \bibinfo {author} {\bibfnamefont {K.~M.}\ \bibnamefont {Awan}}, \bibinfo {author} {\bibfnamefont {S.~A.}\ \bibnamefont {Schulz}}, \bibinfo {author} {\bibfnamefont {J.}~\bibnamefont {Upham}}, \bibinfo {author} {\bibfnamefont {M.~Z.}\ \bibnamefont {Alam}}, \bibinfo {author} {\bibfnamefont {J.~E.}\ \bibnamefont {Sipe}},\ and\ \bibinfo {author} {\bibfnamefont {R.~W.}\ \bibnamefont {Boyd}},\ }\bibfield  {title} {\bibinfo {title} {Weak superradiance in arrays of plasmonic nanoantennas},\ }\href {https://doi.org/10.1103/physreva.100.043814} {\bibfield  {journal} {\bibinfo  {journal} {Phys. Rev. A}\ }\textbf {\bibinfo {volume} {100}},\ \bibinfo {pages} {043814} (\bibinfo {year} {2019})}\BibitemShut {NoStop}%
\bibitem [{\citenamefont {Zhu}\ \emph {et~al.}(2024{\natexlab{b}})\citenamefont {Zhu}, \citenamefont {Boehme}, \citenamefont {Feld}, \citenamefont {Moskalenko}, \citenamefont {Dirin}, \citenamefont {Mahrt}, \citenamefont {St\"{o}ferle}, \citenamefont {Bodnarchuk}, \citenamefont {Efros}, \citenamefont {Sercel}, \citenamefont {Kovalenko},\ and\ \citenamefont {Rainò}}]{Zhu2024_single}%
  \BibitemOpen
  \bibfield  {author} {\bibinfo {author} {\bibfnamefont {C.}~\bibnamefont {Zhu}}, \bibinfo {author} {\bibfnamefont {S.~C.}\ \bibnamefont {Boehme}}, \bibinfo {author} {\bibfnamefont {L.~G.}\ \bibnamefont {Feld}}, \bibinfo {author} {\bibfnamefont {A.}~\bibnamefont {Moskalenko}}, \bibinfo {author} {\bibfnamefont {D.~N.}\ \bibnamefont {Dirin}}, \bibinfo {author} {\bibfnamefont {R.~F.}\ \bibnamefont {Mahrt}}, \bibinfo {author} {\bibfnamefont {T.}~\bibnamefont {St\"{o}ferle}}, \bibinfo {author} {\bibfnamefont {M.~I.}\ \bibnamefont {Bodnarchuk}}, \bibinfo {author} {\bibfnamefont {A.~L.}\ \bibnamefont {Efros}}, \bibinfo {author} {\bibfnamefont {P.~C.}\ \bibnamefont {Sercel}}, \bibinfo {author} {\bibfnamefont {M.~V.}\ \bibnamefont {Kovalenko}},\ and\ \bibinfo {author} {\bibfnamefont {G.}~\bibnamefont {Rainò}},\ }\bibfield  {title} {\bibinfo {title} {Single-photon superradiance in individual caesium lead halide quantum dots},\ }\href {https://doi.org/10.1038/s41586-023-07001-8} {\bibfield  {journal} {\bibinfo
  {journal} {Nature}\ }\textbf {\bibinfo {volume} {626}},\ \bibinfo {pages} {535–541} (\bibinfo {year} {2024}{\natexlab{b}})}\BibitemShut {NoStop}%
\bibitem [{\citenamefont {Stiesdal}\ \emph {et~al.}(2020)\citenamefont {Stiesdal}, \citenamefont {Busche}, \citenamefont {Kumlin}, \citenamefont {Kleinbeck}, \citenamefont {B\"uchler},\ and\ \citenamefont {Hofferberth}}]{Stiesdal2020}%
  \BibitemOpen
  \bibfield  {author} {\bibinfo {author} {\bibfnamefont {N.}~\bibnamefont {Stiesdal}}, \bibinfo {author} {\bibfnamefont {H.}~\bibnamefont {Busche}}, \bibinfo {author} {\bibfnamefont {J.}~\bibnamefont {Kumlin}}, \bibinfo {author} {\bibfnamefont {K.}~\bibnamefont {Kleinbeck}}, \bibinfo {author} {\bibfnamefont {H.~P.}\ \bibnamefont {B\"uchler}},\ and\ \bibinfo {author} {\bibfnamefont {S.}~\bibnamefont {Hofferberth}},\ }\bibfield  {title} {\bibinfo {title} {Observation of collective decay dynamics of a single {R}ydberg superatom},\ }\href {https://doi.org/10.1103/PhysRevResearch.2.043339} {\bibfield  {journal} {\bibinfo  {journal} {Phys. Rev. Res.}\ }\textbf {\bibinfo {volume} {2}},\ \bibinfo {pages} {043339} (\bibinfo {year} {2020})}\BibitemShut {NoStop}%
\bibitem [{\citenamefont {Glicenstein}\ \emph {et~al.}(2022)\citenamefont {Glicenstein}, \citenamefont {Ferioli}, \citenamefont {Browaeys},\ and\ \citenamefont {Ferrier-Barbut}}]{Glicenstein2022}%
  \BibitemOpen
  \bibfield  {author} {\bibinfo {author} {\bibfnamefont {A.}~\bibnamefont {Glicenstein}}, \bibinfo {author} {\bibfnamefont {G.}~\bibnamefont {Ferioli}}, \bibinfo {author} {\bibfnamefont {A.}~\bibnamefont {Browaeys}},\ and\ \bibinfo {author} {\bibfnamefont {I.}~\bibnamefont {Ferrier-Barbut}},\ }\bibfield  {title} {\bibinfo {title} {From superradiance to subradiance: exploring the many-body dicke ladder},\ }\href {https://doi.org/10.1364/OL.451903} {\bibfield  {journal} {\bibinfo  {journal} {Opt. Lett.}\ }\textbf {\bibinfo {volume} {47}},\ \bibinfo {pages} {1541} (\bibinfo {year} {2022})}\BibitemShut {NoStop}%
\bibitem [{\citenamefont {Pennetta}\ \emph {et~al.}(2022)\citenamefont {Pennetta}, \citenamefont {Lechner}, \citenamefont {Blaha}, \citenamefont {Rauschenbeutel}, \citenamefont {Schneeweiss},\ and\ \citenamefont {Volz}}]{Pennetta2022}%
  \BibitemOpen
  \bibfield  {author} {\bibinfo {author} {\bibfnamefont {R.}~\bibnamefont {Pennetta}}, \bibinfo {author} {\bibfnamefont {D.}~\bibnamefont {Lechner}}, \bibinfo {author} {\bibfnamefont {M.}~\bibnamefont {Blaha}}, \bibinfo {author} {\bibfnamefont {A.}~\bibnamefont {Rauschenbeutel}}, \bibinfo {author} {\bibfnamefont {P.}~\bibnamefont {Schneeweiss}},\ and\ \bibinfo {author} {\bibfnamefont {J.}~\bibnamefont {Volz}},\ }\bibfield  {title} {\bibinfo {title} {Observation of {C}oherent {C}oupling between {S}uper- and {S}ubradiant {S}tates of an {E}nsemble of {C}old {A}toms {C}ollectively {C}oupled to a {S}ingle {P}ropagating {O}ptical {M}ode},\ }\href {https://doi.org/10.1103/PhysRevLett.128.203601} {\bibfield  {journal} {\bibinfo  {journal} {Phys. Rev. Lett.}\ }\textbf {\bibinfo {volume} {128}},\ \bibinfo {pages} {203601} (\bibinfo {year} {2022})}\BibitemShut {NoStop}%
\bibitem [{\citenamefont {Żakowicz}(1971)}]{Zakowicz1971}%
  \BibitemOpen
  \bibfield  {author} {\bibinfo {author} {\bibfnamefont {W.}~\bibnamefont {Żakowicz}},\ }\bibfield  {title} {\bibinfo {title} {Superradiant states in two-level and harmonic-oscillator systems},\ }\href {https://doi.org/10.1007/bf02770094} {\bibfield  {journal} {\bibinfo  {journal} {Lett. Nuovo Cimento 2}\ }\textbf {\bibinfo {volume} {2}},\ \bibinfo {pages} {99–103} (\bibinfo {year} {1971})}\BibitemShut {NoStop}%
\bibitem [{\citenamefont {Delanty}\ \emph {et~al.}(2011)\citenamefont {Delanty}, \citenamefont {Rebic},\ and\ \citenamefont {Twamley}}]{delanty2011superradianceharmonicoscillators}%
  \BibitemOpen
  \bibfield  {author} {\bibinfo {author} {\bibfnamefont {M.}~\bibnamefont {Delanty}}, \bibinfo {author} {\bibfnamefont {S.}~\bibnamefont {Rebic}},\ and\ \bibinfo {author} {\bibfnamefont {J.}~\bibnamefont {Twamley}},\ }\href {https://arxiv.org/abs/1107.5080} {\bibinfo {title} {Superradiance of {H}armonic {O}scillators}} (\bibinfo {year} {2011}),\ \Eprint {https://arxiv.org/abs/1107.5080} {arXiv:1107.5080 [quant-ph]} \BibitemShut {NoStop}%
\bibitem [{\citenamefont {Delanty}\ \emph {et~al.}(2012)\citenamefont {Delanty}, \citenamefont {Rebi\'{c}},\ and\ \citenamefont {Twamley}}]{Delanty2012}%
  \BibitemOpen
  \bibfield  {author} {\bibinfo {author} {\bibfnamefont {M.}~\bibnamefont {Delanty}}, \bibinfo {author} {\bibfnamefont {S.}~\bibnamefont {Rebi\'{c}}},\ and\ \bibinfo {author} {\bibfnamefont {J.}~\bibnamefont {Twamley}},\ }\bibfield  {title} {\bibinfo {title} {Novel collective effects in integrated photonics},\ }\href {https://doi.org/10.1140/epjd/e2012-30044-2} {\bibfield  {journal} {\bibinfo  {journal} {Eur. Phys. J. D}\ }\textbf {\bibinfo {volume} {66}},\ \bibinfo {pages} {93} (\bibinfo {year} {2012})}\BibitemShut {NoStop}%
\bibitem [{\citenamefont {Orell}\ \emph {et~al.}(2022)\citenamefont {Orell}, \citenamefont {Zanner}, \citenamefont {Juan}, \citenamefont {Sharafiev}, \citenamefont {Albert}, \citenamefont {Oleschko}, \citenamefont {Kirchmair},\ and\ \citenamefont {Silveri}}]{Orell2022}%
  \BibitemOpen
  \bibfield  {author} {\bibinfo {author} {\bibfnamefont {T.}~\bibnamefont {Orell}}, \bibinfo {author} {\bibfnamefont {M.}~\bibnamefont {Zanner}}, \bibinfo {author} {\bibfnamefont {M.~L.}\ \bibnamefont {Juan}}, \bibinfo {author} {\bibfnamefont {A.}~\bibnamefont {Sharafiev}}, \bibinfo {author} {\bibfnamefont {R.}~\bibnamefont {Albert}}, \bibinfo {author} {\bibfnamefont {S.}~\bibnamefont {Oleschko}}, \bibinfo {author} {\bibfnamefont {G.}~\bibnamefont {Kirchmair}},\ and\ \bibinfo {author} {\bibfnamefont {M.}~\bibnamefont {Silveri}},\ }\bibfield  {title} {\bibinfo {title} {Collective bosonic effects in an array of transmon devices},\ }\href {https://doi.org/10.1103/PhysRevA.105.063701} {\bibfield  {journal} {\bibinfo  {journal} {Phys. Rev. A}\ }\textbf {\bibinfo {volume} {105}},\ \bibinfo {pages} {063701} (\bibinfo {year} {2022})}\BibitemShut {NoStop}%
\bibitem [{\citenamefont {Fox}(2006)}]{Fox2006}%
  \BibitemOpen
  \bibfield  {author} {\bibinfo {author} {\bibfnamefont {M.}~\bibnamefont {Fox}},\ }\href@noop {} {\emph {\bibinfo {title} {Quantum {O}ptics: {A}n {I}ntroduction}}},\ \bibinfo {series} {Oxford Master Series in Physics}\ No.~\bibinfo {number} {15}\ (\bibinfo  {publisher} {Oxford University Press},\ \bibinfo {year} {2006})\BibitemShut {NoStop}%
\bibitem [{\citenamefont {May}\ and\ \citenamefont {K\"{u}hn}(2011)}]{May2011}%
  \BibitemOpen
  \bibfield  {author} {\bibinfo {author} {\bibfnamefont {V.}~\bibnamefont {May}}\ and\ \bibinfo {author} {\bibfnamefont {O.}~\bibnamefont {K\"{u}hn}},\ }\href {https://doi.org/10.1002/9783527633791} {\emph {\bibinfo {title} {Charge and {E}nergy {T}ransfer {D}ynamics in {M}olecular {S}ystems}}}\ (\bibinfo  {publisher} {Wiley},\ \bibinfo {year} {2011})\BibitemShut {NoStop}%
\bibitem [{\citenamefont {Mok}\ \emph {et~al.}(2024)\citenamefont {Mok}, \citenamefont {Poddar}, \citenamefont {Sierra}, \citenamefont {Rusconi}, \citenamefont {Preskill},\ and\ \citenamefont {Asenjo-Garcia}}]{Mok2024}%
  \BibitemOpen
  \bibfield  {author} {\bibinfo {author} {\bibfnamefont {W.}~\bibnamefont {Mok}}, \bibinfo {author} {\bibfnamefont {A.}~\bibnamefont {Poddar}}, \bibinfo {author} {\bibfnamefont {E.}~\bibnamefont {Sierra}}, \bibinfo {author} {\bibfnamefont {C.~C.}\ \bibnamefont {Rusconi}}, \bibinfo {author} {\bibfnamefont {J.}~\bibnamefont {Preskill}},\ and\ \bibinfo {author} {\bibfnamefont {A.}~\bibnamefont {Asenjo-Garcia}},\ }\href {https://arxiv.org/abs/2406.00722} {\bibinfo {title} {Universal scaling laws for correlated decay of many-body quantum systems}} (\bibinfo {year} {2024}),\ \Eprint {https://arxiv.org/abs/2406.00722} {arXiv:2406.00722 [quant-ph]} \BibitemShut {NoStop}%
\bibitem [{\citenamefont {Rosario}\ \emph {et~al.}(2025)\citenamefont {Rosario}, \citenamefont {Solak}, \citenamefont {Cidrim}, \citenamefont {Bachelard},\ and\ \citenamefont {Schachenmayer}}]{Rosario2025}%
  \BibitemOpen
  \bibfield  {author} {\bibinfo {author} {\bibfnamefont {P.}~\bibnamefont {Rosario}}, \bibinfo {author} {\bibfnamefont {L.~O.~R.}\ \bibnamefont {Solak}}, \bibinfo {author} {\bibfnamefont {A.}~\bibnamefont {Cidrim}}, \bibinfo {author} {\bibfnamefont {R.}~\bibnamefont {Bachelard}},\ and\ \bibinfo {author} {\bibfnamefont {J.}~\bibnamefont {Schachenmayer}},\ }\bibfield  {title} {\bibinfo {title} {Unraveling {D}icke {S}uperradiant {D}ecay with {S}eparable {C}oherent {S}pin {S}tates},\ }\href {https://doi.org/10.1103/xcxr-sm9c} {\bibfield  {journal} {\bibinfo  {journal} {Phys. Rev. Lett.}\ }\textbf {\bibinfo {volume} {135}},\ \bibinfo {pages} {133602} (\bibinfo {year} {2025})}\BibitemShut {NoStop}%
\bibitem [{\citenamefont {Spano}\ and\ \citenamefont {Mukamel}(1989)}]{Spano1989}%
  \BibitemOpen
  \bibfield  {author} {\bibinfo {author} {\bibfnamefont {F.~C.}\ \bibnamefont {Spano}}\ and\ \bibinfo {author} {\bibfnamefont {S.}~\bibnamefont {Mukamel}},\ }\bibfield  {title} {\bibinfo {title} {Superradiance in molecular aggregates},\ }\href {https://doi.org/10.1063/1.457174} {\bibfield  {journal} {\bibinfo  {journal} {J. Chem. Phys.}\ }\textbf {\bibinfo {volume} {91}},\ \bibinfo {pages} {683–700} (\bibinfo {year} {1989})}\BibitemShut {NoStop}%
\bibitem [{\citenamefont {Sakurai}\ and\ \citenamefont {Napolitano}(2020)}]{Sakurai2020-di}%
  \BibitemOpen
  \bibfield  {author} {\bibinfo {author} {\bibfnamefont {J.~J.}\ \bibnamefont {Sakurai}}\ and\ \bibinfo {author} {\bibfnamefont {J.}~\bibnamefont {Napolitano}},\ }\href@noop {} {\emph {\bibinfo {title} {Modern quantum mechanics}}},\ \bibinfo {edition} {3rd}\ ed.\ (\bibinfo  {publisher} {Cambridge University Press (Virtual Publishing)},\ \bibinfo {address} {Cambridge, England},\ \bibinfo {year} {2020})\BibitemShut {NoStop}%
\bibitem [{\citenamefont {Crubellier}\ \emph {et~al.}(1980)\citenamefont {Crubellier}, \citenamefont {Liberman},\ and\ \citenamefont {Pillet}}]{Crubellier1980}%
  \BibitemOpen
  \bibfield  {author} {\bibinfo {author} {\bibfnamefont {A.}~\bibnamefont {Crubellier}}, \bibinfo {author} {\bibfnamefont {S.}~\bibnamefont {Liberman}},\ and\ \bibinfo {author} {\bibfnamefont {P.}~\bibnamefont {Pillet}},\ }\bibfield  {title} {\bibinfo {title} {Superradiance and subradiance in three-level systems},\ }\href {https://doi.org/10.1016/0030-4018(80)90181-9} {\bibfield  {journal} {\bibinfo  {journal} {Opt. Commun.}\ }\textbf {\bibinfo {volume} {33}},\ \bibinfo {pages} {143–148} (\bibinfo {year} {1980})}\BibitemShut {NoStop}%
\bibitem [{\citenamefont {Sutherland}\ and\ \citenamefont {Robicheaux}(2017)}]{Sutherland2017}%
  \BibitemOpen
  \bibfield  {author} {\bibinfo {author} {\bibfnamefont {R.~T.}\ \bibnamefont {Sutherland}}\ and\ \bibinfo {author} {\bibfnamefont {F.}~\bibnamefont {Robicheaux}},\ }\bibfield  {title} {\bibinfo {title} {Superradiance in inverted multilevel atomic clouds},\ }\href {https://doi.org/10.1103/PhysRevA.95.033839} {\bibfield  {journal} {\bibinfo  {journal} {Phys. Rev. A}\ }\textbf {\bibinfo {volume} {95}},\ \bibinfo {pages} {033839} (\bibinfo {year} {2017})}\BibitemShut {NoStop}%
\bibitem [{\citenamefont {Fisher}\ \emph {et~al.}(1989)\citenamefont {Fisher}, \citenamefont {Weichman}, \citenamefont {Grinstein},\ and\ \citenamefont {Fisher}}]{Fisher1989}%
  \BibitemOpen
  \bibfield  {author} {\bibinfo {author} {\bibfnamefont {M.~P.~A.}\ \bibnamefont {Fisher}}, \bibinfo {author} {\bibfnamefont {P.~B.}\ \bibnamefont {Weichman}}, \bibinfo {author} {\bibfnamefont {G.}~\bibnamefont {Grinstein}},\ and\ \bibinfo {author} {\bibfnamefont {D.~S.}\ \bibnamefont {Fisher}},\ }\bibfield  {title} {\bibinfo {title} {Boson localization and the superfluid-insulator transition},\ }\href {https://doi.org/10.1103/PhysRevB.40.546} {\bibfield  {journal} {\bibinfo  {journal} {Phys. Rev. B}\ }\textbf {\bibinfo {volume} {40}},\ \bibinfo {pages} {546} (\bibinfo {year} {1989})}\BibitemShut {NoStop}%
\bibitem [{\citenamefont {Roushan}\ \emph {et~al.}(2017)\citenamefont {Roushan}, \citenamefont {Neill}, \citenamefont {Tangpanitanon}, \citenamefont {Bastidas}, \citenamefont {Megrant}, \citenamefont {Barends}, \citenamefont {Chen}, \citenamefont {Chen}, \citenamefont {Chiaro}, \citenamefont {Dunsworth}, \citenamefont {Fowler}, \citenamefont {Foxen}, \citenamefont {Giustina}, \citenamefont {Jeffrey}, \citenamefont {Kelly}, \citenamefont {Lucero}, \citenamefont {Mutus}, \citenamefont {Neeley}, \citenamefont {Quintana}, \citenamefont {Sank}, \citenamefont {Vainsencher}, \citenamefont {Wenner}, \citenamefont {White}, \citenamefont {Neven}, \citenamefont {Angelakis},\ and\ \citenamefont {Martinis}}]{Roushan2017}%
  \BibitemOpen
  \bibfield  {author} {\bibinfo {author} {\bibfnamefont {P.}~\bibnamefont {Roushan}}, \bibinfo {author} {\bibfnamefont {C.}~\bibnamefont {Neill}}, \bibinfo {author} {\bibfnamefont {J.}~\bibnamefont {Tangpanitanon}}, \bibinfo {author} {\bibfnamefont {V.~M.}\ \bibnamefont {Bastidas}}, \bibinfo {author} {\bibfnamefont {A.}~\bibnamefont {Megrant}}, \bibinfo {author} {\bibfnamefont {R.}~\bibnamefont {Barends}}, \bibinfo {author} {\bibfnamefont {Y.}~\bibnamefont {Chen}}, \bibinfo {author} {\bibfnamefont {Z.}~\bibnamefont {Chen}}, \bibinfo {author} {\bibfnamefont {B.}~\bibnamefont {Chiaro}}, \bibinfo {author} {\bibfnamefont {A.}~\bibnamefont {Dunsworth}}, \bibinfo {author} {\bibfnamefont {A.}~\bibnamefont {Fowler}}, \bibinfo {author} {\bibfnamefont {B.}~\bibnamefont {Foxen}}, \bibinfo {author} {\bibfnamefont {M.}~\bibnamefont {Giustina}}, \bibinfo {author} {\bibfnamefont {E.}~\bibnamefont {Jeffrey}}, \bibinfo {author} {\bibfnamefont {J.}~\bibnamefont {Kelly}}, \bibinfo {author} {\bibfnamefont {E.}~\bibnamefont
  {Lucero}}, \bibinfo {author} {\bibfnamefont {J.}~\bibnamefont {Mutus}}, \bibinfo {author} {\bibfnamefont {M.}~\bibnamefont {Neeley}}, \bibinfo {author} {\bibfnamefont {C.}~\bibnamefont {Quintana}}, \bibinfo {author} {\bibfnamefont {D.}~\bibnamefont {Sank}}, \bibinfo {author} {\bibfnamefont {A.}~\bibnamefont {Vainsencher}}, \bibinfo {author} {\bibfnamefont {J.}~\bibnamefont {Wenner}}, \bibinfo {author} {\bibfnamefont {T.}~\bibnamefont {White}}, \bibinfo {author} {\bibfnamefont {H.}~\bibnamefont {Neven}}, \bibinfo {author} {\bibfnamefont {D.~G.}\ \bibnamefont {Angelakis}},\ and\ \bibinfo {author} {\bibfnamefont {J.}~\bibnamefont {Martinis}},\ }\bibfield  {title} {\bibinfo {title} {Spectroscopic signatures of localization with interacting photons in superconducting qubits},\ }\href {https://doi.org/10.1126/science.aao1401} {\bibfield  {journal} {\bibinfo  {journal} {Science}\ }\textbf {\bibinfo {volume} {358}},\ \bibinfo {pages} {1175–1179} (\bibinfo {year} {2017})}\BibitemShut {NoStop}%
\bibitem [{\citenamefont {Celardo}\ \emph {et~al.}(2014{\natexlab{a}})\citenamefont {Celardo}, \citenamefont {Giusteri},\ and\ \citenamefont {Borgonovi}}]{Celardo2014_static}%
  \BibitemOpen
  \bibfield  {author} {\bibinfo {author} {\bibfnamefont {G.~L.}\ \bibnamefont {Celardo}}, \bibinfo {author} {\bibfnamefont {G.~G.}\ \bibnamefont {Giusteri}},\ and\ \bibinfo {author} {\bibfnamefont {F.}~\bibnamefont {Borgonovi}},\ }\bibfield  {title} {\bibinfo {title} {Cooperative robustness to static disorder: {S}uperradiance and localization in a nanoscale ring to model light-harvesting systems found in nature},\ }\href {https://doi.org/10.1103/physrevb.90.075113} {\bibfield  {journal} {\bibinfo  {journal} {Phys. Rev. B}\ }\textbf {\bibinfo {volume} {90}},\ \bibinfo {pages} {075113} (\bibinfo {year} {2014}{\natexlab{a}})}\BibitemShut {NoStop}%
\bibitem [{\citenamefont {Celardo}\ \emph {et~al.}(2014{\natexlab{b}})\citenamefont {Celardo}, \citenamefont {Poli}, \citenamefont {Lussardi},\ and\ \citenamefont {Borgonovi}}]{Celardo2014_dynamic}%
  \BibitemOpen
  \bibfield  {author} {\bibinfo {author} {\bibfnamefont {G.~L.}\ \bibnamefont {Celardo}}, \bibinfo {author} {\bibfnamefont {P.}~\bibnamefont {Poli}}, \bibinfo {author} {\bibfnamefont {L.}~\bibnamefont {Lussardi}},\ and\ \bibinfo {author} {\bibfnamefont {F.}~\bibnamefont {Borgonovi}},\ }\bibfield  {title} {\bibinfo {title} {Cooperative robustness to dephasing: {S}ingle-exciton superradiance in a nanoscale ring to model natural light-harvesting systems},\ }\href {https://doi.org/10.1103/physrevb.90.085142} {\bibfield  {journal} {\bibinfo  {journal} {Phys. Rev. B}\ }\textbf {\bibinfo {volume} {90}},\ \bibinfo {pages} {085142} (\bibinfo {year} {2014}{\natexlab{b}})}\BibitemShut {NoStop}%
\bibitem [{\citenamefont {Chen}\ \emph {et~al.}(2022)\citenamefont {Chen}, \citenamefont {Zhou}, \citenamefont {Sukharev}, \citenamefont {Subotnik},\ and\ \citenamefont {Nitzan}}]{Chen2022}%
  \BibitemOpen
  \bibfield  {author} {\bibinfo {author} {\bibfnamefont {H.-T.}\ \bibnamefont {Chen}}, \bibinfo {author} {\bibfnamefont {Z.}~\bibnamefont {Zhou}}, \bibinfo {author} {\bibfnamefont {M.}~\bibnamefont {Sukharev}}, \bibinfo {author} {\bibfnamefont {J.~E.}\ \bibnamefont {Subotnik}},\ and\ \bibinfo {author} {\bibfnamefont {A.}~\bibnamefont {Nitzan}},\ }\bibfield  {title} {\bibinfo {title} {Interplay between disorder and collective coherent response: {S}uperradiance and spectral motional narrowing in the time domain},\ }\href {https://doi.org/10.1103/PhysRevA.106.053703} {\bibfield  {journal} {\bibinfo  {journal} {Phys. Rev. A}\ }\textbf {\bibinfo {volume} {106}},\ \bibinfo {pages} {053703} (\bibinfo {year} {2022})}\BibitemShut {NoStop}%
\bibitem [{\citenamefont {Miftasani}\ and\ \citenamefont {Machnikowski}(2016)}]{Miftasani2016}%
  \BibitemOpen
  \bibfield  {author} {\bibinfo {author} {\bibfnamefont {F.}~\bibnamefont {Miftasani}}\ and\ \bibinfo {author} {\bibfnamefont {P.}~\bibnamefont {Machnikowski}},\ }\bibfield  {title} {\bibinfo {title} {Photon-photon correlation statistics in the collective emission from ensembles of self-assembled quantum dots},\ }\href {https://doi.org/10.1103/PhysRevB.93.075311} {\bibfield  {journal} {\bibinfo  {journal} {Phys. Rev. B}\ }\textbf {\bibinfo {volume} {93}},\ \bibinfo {pages} {075311} (\bibinfo {year} {2016})}\BibitemShut {NoStop}%
\bibitem [{\citenamefont {Blach}\ \emph {et~al.}(2022)\citenamefont {Blach}, \citenamefont {Lumsargis}, \citenamefont {Clark}, \citenamefont {Chuang}, \citenamefont {Wang}, \citenamefont {Dou}, \citenamefont {Schaller}, \citenamefont {Cao}, \citenamefont {Li},\ and\ \citenamefont {Huang}}]{Blach2022}%
  \BibitemOpen
  \bibfield  {author} {\bibinfo {author} {\bibfnamefont {D.~D.}\ \bibnamefont {Blach}}, \bibinfo {author} {\bibfnamefont {V.~A.}\ \bibnamefont {Lumsargis}}, \bibinfo {author} {\bibfnamefont {D.~E.}\ \bibnamefont {Clark}}, \bibinfo {author} {\bibfnamefont {C.}~\bibnamefont {Chuang}}, \bibinfo {author} {\bibfnamefont {K.}~\bibnamefont {Wang}}, \bibinfo {author} {\bibfnamefont {L.}~\bibnamefont {Dou}}, \bibinfo {author} {\bibfnamefont {R.~D.}\ \bibnamefont {Schaller}}, \bibinfo {author} {\bibfnamefont {J.}~\bibnamefont {Cao}}, \bibinfo {author} {\bibfnamefont {C.~W.}\ \bibnamefont {Li}},\ and\ \bibinfo {author} {\bibfnamefont {L.}~\bibnamefont {Huang}},\ }\bibfield  {title} {\bibinfo {title} {Superradiance and {E}xciton {D}elocalization in {P}erovskite {Q}uantum {D}ot {S}uperlattices},\ }\href {https://doi.org/10.1021/acs.nanolett.2c02427} {\bibfield  {journal} {\bibinfo  {journal} {Nano Lett.}\ }\textbf {\bibinfo {volume} {22}},\ \bibinfo {pages} {7811–7818} (\bibinfo {year} {2022})}\BibitemShut {NoStop}%
\bibitem [{\citenamefont {Luo}\ \emph {et~al.}(2025)\citenamefont {Luo}, \citenamefont {Tang}, \citenamefont {Park}, \citenamefont {Wang}, \citenamefont {Park}, \citenamefont {Khurana}, \citenamefont {Singh}, \citenamefont {Cheon}, \citenamefont {Belyanin}, \citenamefont {Sokolov},\ and\ \citenamefont {Son}}]{Luo2025}%
  \BibitemOpen
  \bibfield  {author} {\bibinfo {author} {\bibfnamefont {L.}~\bibnamefont {Luo}}, \bibinfo {author} {\bibfnamefont {X.}~\bibnamefont {Tang}}, \bibinfo {author} {\bibfnamefont {J.}~\bibnamefont {Park}}, \bibinfo {author} {\bibfnamefont {C.-W.}\ \bibnamefont {Wang}}, \bibinfo {author} {\bibfnamefont {M.}~\bibnamefont {Park}}, \bibinfo {author} {\bibfnamefont {M.}~\bibnamefont {Khurana}}, \bibinfo {author} {\bibfnamefont {A.}~\bibnamefont {Singh}}, \bibinfo {author} {\bibfnamefont {J.}~\bibnamefont {Cheon}}, \bibinfo {author} {\bibfnamefont {A.}~\bibnamefont {Belyanin}}, \bibinfo {author} {\bibfnamefont {A.~V.}\ \bibnamefont {Sokolov}},\ and\ \bibinfo {author} {\bibfnamefont {D.~H.}\ \bibnamefont {Son}},\ }\bibfield  {title} {\bibinfo {title} {Polarized {S}uperradiance from {C}s{P}b{B}r3 {Q}uantum {D}ot {S}uperlattice with {C}ontrolled {I}nterdot {E}lectronic {C}oupling},\ }\href {https://doi.org/10.1021/acs.nanolett.5c00478} {\bibfield  {journal} {\bibinfo  {journal} {Nano Lett.}\ }\textbf {\bibinfo {volume}
  {25}},\ \bibinfo {pages} {6176–6183} (\bibinfo {year} {2025})}\BibitemShut {NoStop}%
\bibitem [{\citenamefont {Potma}\ and\ \citenamefont {Wiersma}(1998)}]{Potma1998}%
  \BibitemOpen
  \bibfield  {author} {\bibinfo {author} {\bibfnamefont {E.~O.}\ \bibnamefont {Potma}}\ and\ \bibinfo {author} {\bibfnamefont {D.~A.}\ \bibnamefont {Wiersma}},\ }\bibfield  {title} {\bibinfo {title} {Exciton superradiance in aggregates: {T}he effect of disorder, higher order exciton-phonon coupling and dimensionality},\ }\href {https://doi.org/10.1063/1.475898} {\bibfield  {journal} {\bibinfo  {journal} {J. Chem. Phys.}\ }\textbf {\bibinfo {volume} {108}},\ \bibinfo {pages} {4894–4903} (\bibinfo {year} {1998})}\BibitemShut {NoStop}%
\bibitem [{\citenamefont {Celardo}\ \emph {et~al.}(2013)\citenamefont {Celardo}, \citenamefont {Biella}, \citenamefont {Kaplan},\ and\ \citenamefont {Borgonovi}}]{Celardo2013}%
  \BibitemOpen
  \bibfield  {author} {\bibinfo {author} {\bibfnamefont {G.~L.}\ \bibnamefont {Celardo}}, \bibinfo {author} {\bibfnamefont {A.}~\bibnamefont {Biella}}, \bibinfo {author} {\bibfnamefont {L.}~\bibnamefont {Kaplan}},\ and\ \bibinfo {author} {\bibfnamefont {F.}~\bibnamefont {Borgonovi}},\ }\bibfield  {title} {\bibinfo {title} {Interplay of superradiance and disorder in the {A}nderson model},\ }\href {https://doi.org/10.1002/prop.201200082} {\bibfield  {journal} {\bibinfo  {journal} {Fortschr. Phys.}\ }\textbf {\bibinfo {volume} {61}},\ \bibinfo {pages} {250} (\bibinfo {year} {2013})}\BibitemShut {NoStop}%
\bibitem [{\citenamefont {Smyth}\ \emph {et~al.}(2015)\citenamefont {Smyth}, \citenamefont {Oblinsky},\ and\ \citenamefont {Scholes}}]{Smyth2015}%
  \BibitemOpen
  \bibfield  {author} {\bibinfo {author} {\bibfnamefont {C.}~\bibnamefont {Smyth}}, \bibinfo {author} {\bibfnamefont {D.~G.}\ \bibnamefont {Oblinsky}},\ and\ \bibinfo {author} {\bibfnamefont {G.~D.}\ \bibnamefont {Scholes}},\ }\bibfield  {title} {\bibinfo {title} {B800–{B}850 coherence correlates with energy transfer rates in the {L}{H}2 complex of photosynthetic purple bacteria},\ }\href {https://doi.org/10.1039/c5cp00295h} {\bibfield  {journal} {\bibinfo  {journal} {Phys. Chem. Chem. Phys.}\ }\textbf {\bibinfo {volume} {17}},\ \bibinfo {pages} {30805–30816} (\bibinfo {year} {2015})}\BibitemShut {NoStop}%
\bibitem [{\citenamefont {Balzer}\ and\ \citenamefont {Kassal}(2022)}]{Balzer2022}%
  \BibitemOpen
  \bibfield  {author} {\bibinfo {author} {\bibfnamefont {D.}~\bibnamefont {Balzer}}\ and\ \bibinfo {author} {\bibfnamefont {I.}~\bibnamefont {Kassal}},\ }\bibfield  {title} {\bibinfo {title} {Even a little delocalization produces large kinetic enhancements of charge-separation efficiency in organic photovoltaics},\ }\href {https://doi.org/10.1126/sciadv.abl9692} {\bibfield  {journal} {\bibinfo  {journal} {Sci. Adv.}\ }\textbf {\bibinfo {volume} {8}},\ \bibinfo {pages} {eabl9692} (\bibinfo {year} {2022})}\BibitemShut {NoStop}%
\bibitem [{\citenamefont {Balzer}\ and\ \citenamefont {Kassal}(2023)}]{Balzer2023}%
  \BibitemOpen
  \bibfield  {author} {\bibinfo {author} {\bibfnamefont {D.}~\bibnamefont {Balzer}}\ and\ \bibinfo {author} {\bibfnamefont {I.}~\bibnamefont {Kassal}},\ }\bibfield  {title} {\bibinfo {title} {Mechanism of {D}elocalization-{E}nhanced {E}xciton {T}ransport in {D}isordered {O}rganic {S}emiconductors},\ }\href {https://doi.org/10.1021/acs.jpclett.2c03886} {\bibfield  {journal} {\bibinfo  {journal} {J. Phys. Chem. Lett.}\ }\textbf {\bibinfo {volume} {14}},\ \bibinfo {pages} {2155–2162} (\bibinfo {year} {2023})}\BibitemShut {NoStop}%
\bibitem [{\citenamefont {Balzer}\ and\ \citenamefont {Kassal}(2024)}]{Balzer2024}%
  \BibitemOpen
  \bibfield  {author} {\bibinfo {author} {\bibfnamefont {D.}~\bibnamefont {Balzer}}\ and\ \bibinfo {author} {\bibfnamefont {I.}~\bibnamefont {Kassal}},\ }\bibfield  {title} {\bibinfo {title} {Delocalisation enables efficient charge generation in organic photovoltaics, even with little to no energetic offset},\ }\href {https://doi.org/10.1039/d3sc05409h} {\bibfield  {journal} {\bibinfo  {journal} {Chem. Sci.}\ }\textbf {\bibinfo {volume} {15}},\ \bibinfo {pages} {4779–4789} (\bibinfo {year} {2024})}\BibitemShut {NoStop}%
\bibitem [{\citenamefont {Ribeiro}\ \emph {et~al.}(2018)\citenamefont {Ribeiro}, \citenamefont {Martínez-Martínez}, \citenamefont {Du}, \citenamefont {Campos-Gonzalez-Angulo},\ and\ \citenamefont {Yuen-Zhou}}]{Ribeiro2018}%
  \BibitemOpen
  \bibfield  {author} {\bibinfo {author} {\bibfnamefont {R.~F.}\ \bibnamefont {Ribeiro}}, \bibinfo {author} {\bibfnamefont {L.~A.}\ \bibnamefont {Martínez-Martínez}}, \bibinfo {author} {\bibfnamefont {M.}~\bibnamefont {Du}}, \bibinfo {author} {\bibfnamefont {J.}~\bibnamefont {Campos-Gonzalez-Angulo}},\ and\ \bibinfo {author} {\bibfnamefont {J.}~\bibnamefont {Yuen-Zhou}},\ }\bibfield  {title} {\bibinfo {title} {Polariton chemistry: controlling molecular dynamics with optical cavities},\ }\href {https://doi.org/10.1039/c8sc01043a} {\bibfield  {journal} {\bibinfo  {journal} {Chem. Sci.}\ }\textbf {\bibinfo {volume} {9}},\ \bibinfo {pages} {6325–6339} (\bibinfo {year} {2018})}\BibitemShut {NoStop}%
\bibitem [{\citenamefont {Engelhardt}\ and\ \citenamefont {Cao}(2023)}]{Engelhardt2023}%
  \BibitemOpen
  \bibfield  {author} {\bibinfo {author} {\bibfnamefont {G.}~\bibnamefont {Engelhardt}}\ and\ \bibinfo {author} {\bibfnamefont {J.}~\bibnamefont {Cao}},\ }\bibfield  {title} {\bibinfo {title} {Polariton {L}ocalization and {D}ispersion {P}roperties of {D}isordered {Q}uantum {E}mitters in {M}ultimode {M}icrocavities},\ }\href {https://doi.org/10.1103/PhysRevLett.130.213602} {\bibfield  {journal} {\bibinfo  {journal} {Phys. Rev. Lett.}\ }\textbf {\bibinfo {volume} {130}},\ \bibinfo {pages} {213602} (\bibinfo {year} {2023})}\BibitemShut {NoStop}%
\bibitem [{\citenamefont {P\'erez-S\'anchez}\ \emph {et~al.}(2024)\citenamefont {P\'erez-S\'anchez}, \citenamefont {Mellini}, \citenamefont {Giebink},\ and\ \citenamefont {Yuen-Zhou}}]{Sanchez2024}%
  \BibitemOpen
  \bibfield  {author} {\bibinfo {author} {\bibfnamefont {J.~B.}\ \bibnamefont {P\'erez-S\'anchez}}, \bibinfo {author} {\bibfnamefont {F.}~\bibnamefont {Mellini}}, \bibinfo {author} {\bibfnamefont {N.~C.}\ \bibnamefont {Giebink}},\ and\ \bibinfo {author} {\bibfnamefont {J.}~\bibnamefont {Yuen-Zhou}},\ }\bibfield  {title} {\bibinfo {title} {Collective polaritonic effects on chemical dynamics suppressed by disorder},\ }\href {https://doi.org/10.1103/PhysRevResearch.6.013222} {\bibfield  {journal} {\bibinfo  {journal} {Phys. Rev. Res.}\ }\textbf {\bibinfo {volume} {6}},\ \bibinfo {pages} {013222} (\bibinfo {year} {2024})}\BibitemShut {NoStop}%
\bibitem [{\citenamefont {Liu}\ \emph {et~al.}(2025)\citenamefont {Liu}, \citenamefont {Yin},\ and\ \citenamefont {Xiong}}]{Liu2025}%
  \BibitemOpen
  \bibfield  {author} {\bibinfo {author} {\bibfnamefont {T.}~\bibnamefont {Liu}}, \bibinfo {author} {\bibfnamefont {G.}~\bibnamefont {Yin}},\ and\ \bibinfo {author} {\bibfnamefont {W.}~\bibnamefont {Xiong}},\ }\bibfield  {title} {\bibinfo {title} {Unlocking delocalization: how much coupling strength is required to overcome energy disorder in molecular polaritons?},\ }\href {https://doi.org/10.1039/d4sc07053d} {\bibfield  {journal} {\bibinfo  {journal} {Chem. Sci.}\ }\textbf {\bibinfo {volume} {16}},\ \bibinfo {pages} {4676–4683} (\bibinfo {year} {2025})}\BibitemShut {NoStop}%
\bibitem [{\citenamefont {Uhlenbeck}\ and\ \citenamefont {Ornstein}(1930)}]{Uhlenbeck1930}%
  \BibitemOpen
  \bibfield  {author} {\bibinfo {author} {\bibfnamefont {G.~E.}\ \bibnamefont {Uhlenbeck}}\ and\ \bibinfo {author} {\bibfnamefont {L.~S.}\ \bibnamefont {Ornstein}},\ }\bibfield  {title} {\bibinfo {title} {On the theory of the brownian motion},\ }\href {https://doi.org/10.1103/PhysRev.36.823} {\bibfield  {journal} {\bibinfo  {journal} {Phys. Rev.}\ }\textbf {\bibinfo {volume} {36}},\ \bibinfo {pages} {823} (\bibinfo {year} {1930})}\BibitemShut {NoStop}%
\bibitem [{\citenamefont {Gardiner}(1985)}]{Gardiner1985}%
  \BibitemOpen
  \bibfield  {author} {\bibinfo {author} {\bibfnamefont {C.}~\bibnamefont {Gardiner}},\ }\href@noop {} {\emph {\bibinfo {title} {Handbook of {S}tochastic {M}ethods for {P}hysics, {C}hemistry, and the {N}atural {S}ciences}}}\ (\bibinfo  {publisher} {Springer},\ \bibinfo {year} {1985})\BibitemShut {NoStop}%
\bibitem [{\citenamefont {Schuurmans}\ \emph {et~al.}(1982)\citenamefont {Schuurmans}, \citenamefont {Vrehen}, \citenamefont {Polder},\ and\ \citenamefont {Gibbs}}]{Schuurmans1982}%
  \BibitemOpen
  \bibfield  {author} {\bibinfo {author} {\bibfnamefont {M.}~\bibnamefont {Schuurmans}}, \bibinfo {author} {\bibfnamefont {Q.}~\bibnamefont {Vrehen}}, \bibinfo {author} {\bibfnamefont {D.}~\bibnamefont {Polder}},\ and\ \bibinfo {author} {\bibfnamefont {H.}~\bibnamefont {Gibbs}},\ }\bibinfo {title} {Superfluorescence},\ in\ \href {https://doi.org/10.1016/s0065-2199(08)60069-x} {\emph {\bibinfo {booktitle} {Advances in Atomic and Molecular Physics Volume 17}}}\ (\bibinfo  {publisher} {Elsevier},\ \bibinfo {year} {1982})\ p.\ \bibinfo {pages} {167–228}\BibitemShut {NoStop}%
\bibitem [{\citenamefont {Shahbazyan}\ \emph {et~al.}(2000)\citenamefont {Shahbazyan}, \citenamefont {Raikh},\ and\ \citenamefont {Vardeny}}]{Shahbazyan2000}%
  \BibitemOpen
  \bibfield  {author} {\bibinfo {author} {\bibfnamefont {T.~V.}\ \bibnamefont {Shahbazyan}}, \bibinfo {author} {\bibfnamefont {M.~E.}\ \bibnamefont {Raikh}},\ and\ \bibinfo {author} {\bibfnamefont {Z.~V.}\ \bibnamefont {Vardeny}},\ }\bibfield  {title} {\bibinfo {title} {Mesoscopic cooperative emission from a disordered system},\ }\href {https://doi.org/10.1103/PhysRevB.61.13266} {\bibfield  {journal} {\bibinfo  {journal} {Phys. Rev. B}\ }\textbf {\bibinfo {volume} {61}},\ \bibinfo {pages} {13266} (\bibinfo {year} {2000})}\BibitemShut {NoStop}%
\bibitem [{\citenamefont {Wiercinski}\ \emph {et~al.}(2024)\citenamefont {Wiercinski}, \citenamefont {Cygorek},\ and\ \citenamefont {Gauger}}]{Wiercinski2024}%
  \BibitemOpen
  \bibfield  {author} {\bibinfo {author} {\bibfnamefont {J.}~\bibnamefont {Wiercinski}}, \bibinfo {author} {\bibfnamefont {M.}~\bibnamefont {Cygorek}},\ and\ \bibinfo {author} {\bibfnamefont {E.~M.}\ \bibnamefont {Gauger}},\ }\bibfield  {title} {\bibinfo {title} {Role of polaron dressing in superradiant emission dynamics},\ }\href {https://doi.org/10.1103/PhysRevResearch.6.033231} {\bibfield  {journal} {\bibinfo  {journal} {Phys. Rev. Res.}\ }\textbf {\bibinfo {volume} {6}},\ \bibinfo {pages} {033231} (\bibinfo {year} {2024})}\BibitemShut {NoStop}%
\bibitem [{\citenamefont {Tao}\ \emph {et~al.}(2025)\citenamefont {Tao}, \citenamefont {Philbin},\ and\ \citenamefont {Narang}}]{Tao2025}%
  \BibitemOpen
  \bibfield  {author} {\bibinfo {author} {\bibfnamefont {X.}~\bibnamefont {Tao}}, \bibinfo {author} {\bibfnamefont {J.~P.}\ \bibnamefont {Philbin}},\ and\ \bibinfo {author} {\bibfnamefont {P.}~\bibnamefont {Narang}},\ }\bibfield  {title} {\bibinfo {title} {Electronic superradiance mediated by nuclear dynamics},\ }\href {https://doi.org/10.1103/PhysRevResearch.7.013133} {\bibfield  {journal} {\bibinfo  {journal} {Phys. Rev. Res.}\ }\textbf {\bibinfo {volume} {7}},\ \bibinfo {pages} {013133} (\bibinfo {year} {2025})}\BibitemShut {NoStop}%
\bibitem [{\citenamefont {Pallmann}\ \emph {et~al.}(2024)\citenamefont {Pallmann}, \citenamefont {K\"oster}, \citenamefont {Zhang}, \citenamefont {Heupel}, \citenamefont {Eichhorn}, \citenamefont {Popov}, \citenamefont {M\o{}lmer},\ and\ \citenamefont {Hunger}}]{Pallmann2024}%
  \BibitemOpen
  \bibfield  {author} {\bibinfo {author} {\bibfnamefont {M.}~\bibnamefont {Pallmann}}, \bibinfo {author} {\bibfnamefont {K.}~\bibnamefont {K\"oster}}, \bibinfo {author} {\bibfnamefont {Y.}~\bibnamefont {Zhang}}, \bibinfo {author} {\bibfnamefont {J.}~\bibnamefont {Heupel}}, \bibinfo {author} {\bibfnamefont {T.}~\bibnamefont {Eichhorn}}, \bibinfo {author} {\bibfnamefont {C.}~\bibnamefont {Popov}}, \bibinfo {author} {\bibfnamefont {K.}~\bibnamefont {M\o{}lmer}},\ and\ \bibinfo {author} {\bibfnamefont {D.}~\bibnamefont {Hunger}},\ }\bibfield  {title} {\bibinfo {title} {Cavity-{M}ediated {C}ollective {E}mission from {F}ew {E}mitters in a {D}iamond {M}embrane},\ }\href {https://doi.org/10.1103/PhysRevX.14.041055} {\bibfield  {journal} {\bibinfo  {journal} {Phys. Rev. X}\ }\textbf {\bibinfo {volume} {14}},\ \bibinfo {pages} {041055} (\bibinfo {year} {2024})}\BibitemShut {NoStop}%
\bibitem [{\citenamefont {Petrov}\ \emph {et~al.}(2002)\citenamefont {Petrov}, \citenamefont {Zelinskyy},\ and\ \citenamefont {May}}]{Petrov2002}%
  \BibitemOpen
  \bibfield  {author} {\bibinfo {author} {\bibfnamefont {E.~G.}\ \bibnamefont {Petrov}}, \bibinfo {author} {\bibfnamefont {{\relax Ya}.~R.}\ \bibnamefont {Zelinskyy}},\ and\ \bibinfo {author} {\bibfnamefont {V.}~\bibnamefont {May}},\ }\bibfield  {title} {\bibinfo {title} {Bridge {M}ediated {E}lectron {T}ransfer: {A} {U}nified {D}escription of the {T}hermally {A}ctivated and {S}uperexchange {M}echanisms},\ }\href {https://doi.org/10.1021/jp013427g} {\bibfield  {journal} {\bibinfo  {journal} {J. Phys. Chem. B}\ }\textbf {\bibinfo {volume} {106}},\ \bibinfo {pages} {3092} (\bibinfo {year} {2002})}\BibitemShut {NoStop}%
\bibitem [{\citenamefont {Nitzan}(2006)}]{Nitzan2006}%
  \BibitemOpen
  \bibfield  {author} {\bibinfo {author} {\bibfnamefont {A.}~\bibnamefont {Nitzan}},\ }\href {https://doi.org/10.1093/oso/9780198529798.001.0001} {\emph {\bibinfo {title} {Chemical {D}ynamics in {C}ondensed {P}hases: {R}elaxation, {T}ransfer and {R}eactions in {C}ondensed {M}olecular {S}ystems}}}\ (\bibinfo  {publisher} {Oxford University Press},\ \bibinfo {year} {2006})\BibitemShut {NoStop}%
\bibitem [{\citenamefont {Brown}\ and\ \citenamefont {Gauger}(2019)}]{Brown2019}%
  \BibitemOpen
  \bibfield  {author} {\bibinfo {author} {\bibfnamefont {W.~M.}\ \bibnamefont {Brown}}\ and\ \bibinfo {author} {\bibfnamefont {E.~M.}\ \bibnamefont {Gauger}},\ }\bibfield  {title} {\bibinfo {title} {Light {H}arvesting with {G}uide-{S}lide {S}uperabsorbing {C}ondensed-{M}atter {N}anostructures},\ }\href {https://doi.org/10.1021/acs.jpclett.9b01349} {\bibfield  {journal} {\bibinfo  {journal} {J. Phys. Chem. Lett.}\ }\textbf {\bibinfo {volume} {10}},\ \bibinfo {pages} {4323–4329} (\bibinfo {year} {2019})}\BibitemShut {NoStop}%
\bibitem [{\citenamefont {Burgess}\ \emph {et~al.}(2025)\citenamefont {Burgess}, \citenamefont {Waller}, \citenamefont {Gauger},\ and\ \citenamefont {Bennett}}]{Burgess2025}%
  \BibitemOpen
  \bibfield  {author} {\bibinfo {author} {\bibfnamefont {A.}~\bibnamefont {Burgess}}, \bibinfo {author} {\bibfnamefont {M.~C.}\ \bibnamefont {Waller}}, \bibinfo {author} {\bibfnamefont {E.~M.}\ \bibnamefont {Gauger}},\ and\ \bibinfo {author} {\bibfnamefont {R.}~\bibnamefont {Bennett}},\ }\bibfield  {title} {\bibinfo {title} {Engineering {D}ipole-{D}ipole {C}ouplings for {E}nhanced {C}ooperative {L}ight-{M}atter {I}nteractions},\ }\href {https://doi.org/10.1103/PhysRevLett.134.113602} {\bibfield  {journal} {\bibinfo  {journal} {Phys. Rev. Lett.}\ }\textbf {\bibinfo {volume} {134}},\ \bibinfo {pages} {113602} (\bibinfo {year} {2025})}\BibitemShut {NoStop}%
\bibitem [{\citenamefont {Dias}\ \emph {et~al.}(2021)\citenamefont {Dias}, \citenamefont {W\"{a}chtler}, \citenamefont {Bastidas}, \citenamefont {Nemoto},\ and\ \citenamefont {Munro}}]{Dias2021}%
  \BibitemOpen
  \bibfield  {author} {\bibinfo {author} {\bibfnamefont {J.}~\bibnamefont {Dias}}, \bibinfo {author} {\bibfnamefont {C.~W.}\ \bibnamefont {W\"{a}chtler}}, \bibinfo {author} {\bibfnamefont {V.~M.}\ \bibnamefont {Bastidas}}, \bibinfo {author} {\bibfnamefont {K.}~\bibnamefont {Nemoto}},\ and\ \bibinfo {author} {\bibfnamefont {W.~J.}\ \bibnamefont {Munro}},\ }\bibfield  {title} {\bibinfo {title} {Reservoir-assisted energy migration through multiple spin domains},\ }\href {https://doi.org/10.1103/physrevb.104.l140303} {\bibfield  {journal} {\bibinfo  {journal} {Phys. Rev. B}\ }\textbf {\bibinfo {volume} {104}},\ \bibinfo {pages} {L140303} (\bibinfo {year} {2021})}\BibitemShut {NoStop}%
\bibitem [{\citenamefont {Du}\ \emph {et~al.}(2018)\citenamefont {Du}, \citenamefont {Mart\'{\i}nez-Mart\'{\i}nez}, \citenamefont {Ribeiro}, \citenamefont {Hu}, \citenamefont {Menon},\ and\ \citenamefont {Yuen-Zhou}}]{Du2018}%
  \BibitemOpen
  \bibfield  {author} {\bibinfo {author} {\bibfnamefont {M.}~\bibnamefont {Du}}, \bibinfo {author} {\bibfnamefont {L.~A.}\ \bibnamefont {Mart\'{\i}nez-Mart\'{\i}nez}}, \bibinfo {author} {\bibfnamefont {R.~F.}\ \bibnamefont {Ribeiro}}, \bibinfo {author} {\bibfnamefont {Z.}~\bibnamefont {Hu}}, \bibinfo {author} {\bibfnamefont {V.~M.}\ \bibnamefont {Menon}},\ and\ \bibinfo {author} {\bibfnamefont {J.}~\bibnamefont {Yuen-Zhou}},\ }\bibfield  {title} {\bibinfo {title} {Theory for polariton-assisted remote energy transfer},\ }\href {https://doi.org/10.1039/C8SC00171E} {\bibfield  {journal} {\bibinfo  {journal} {Chem. Sci.}\ }\textbf {\bibinfo {volume} {9}},\ \bibinfo {pages} {6659} (\bibinfo {year} {2018})}\BibitemShut {NoStop}%
\bibitem [{\citenamefont {Kaluzny}\ \emph {et~al.}(1983)\citenamefont {Kaluzny}, \citenamefont {Goy}, \citenamefont {Gross}, \citenamefont {Raimond},\ and\ \citenamefont {Haroche}}]{Kaluzny1983}%
  \BibitemOpen
  \bibfield  {author} {\bibinfo {author} {\bibfnamefont {Y.}~\bibnamefont {Kaluzny}}, \bibinfo {author} {\bibfnamefont {P.}~\bibnamefont {Goy}}, \bibinfo {author} {\bibfnamefont {M.}~\bibnamefont {Gross}}, \bibinfo {author} {\bibfnamefont {J.~M.}\ \bibnamefont {Raimond}},\ and\ \bibinfo {author} {\bibfnamefont {S.}~\bibnamefont {Haroche}},\ }\bibfield  {title} {\bibinfo {title} {Observation of self-induced rabi oscillations in two-level atoms excited inside a resonant cavity: The ringing regime of superradiance},\ }\href {https://doi.org/10.1103/PhysRevLett.51.1175} {\bibfield  {journal} {\bibinfo  {journal} {Phys. Rev. Lett.}\ }\textbf {\bibinfo {volume} {51}},\ \bibinfo {pages} {1175} (\bibinfo {year} {1983})}\BibitemShut {NoStop}%
\bibitem [{\citenamefont {Bohnet}\ \emph {et~al.}(2012)\citenamefont {Bohnet}, \citenamefont {Chen}, \citenamefont {Weiner}, \citenamefont {Meiser}, \citenamefont {Holland},\ and\ \citenamefont {Thompson}}]{Bohnet2012}%
  \BibitemOpen
  \bibfield  {author} {\bibinfo {author} {\bibfnamefont {J.~G.}\ \bibnamefont {Bohnet}}, \bibinfo {author} {\bibfnamefont {Z.}~\bibnamefont {Chen}}, \bibinfo {author} {\bibfnamefont {J.~M.}\ \bibnamefont {Weiner}}, \bibinfo {author} {\bibfnamefont {D.}~\bibnamefont {Meiser}}, \bibinfo {author} {\bibfnamefont {M.~J.}\ \bibnamefont {Holland}},\ and\ \bibinfo {author} {\bibfnamefont {J.~K.}\ \bibnamefont {Thompson}},\ }\bibfield  {title} {\bibinfo {title} {A steady-state superradiant laser with less than one intracavity photon},\ }\href {https://doi.org/10.1038/nature10920} {\bibfield  {journal} {\bibinfo  {journal} {Nature}\ }\textbf {\bibinfo {volume} {484}},\ \bibinfo {pages} {78–81} (\bibinfo {year} {2012})}\BibitemShut {NoStop}%
\bibitem [{\citenamefont {Zhou}\ \emph {et~al.}(2015)\citenamefont {Zhou}, \citenamefont {Yi}, \citenamefont {Luk}, \citenamefont {Gan}, \citenamefont {Fan},\ and\ \citenamefont {Yu}}]{Zhou2015}%
  \BibitemOpen
  \bibfield  {author} {\bibinfo {author} {\bibfnamefont {M.}~\bibnamefont {Zhou}}, \bibinfo {author} {\bibfnamefont {S.}~\bibnamefont {Yi}}, \bibinfo {author} {\bibfnamefont {T.~S.}\ \bibnamefont {Luk}}, \bibinfo {author} {\bibfnamefont {Q.}~\bibnamefont {Gan}}, \bibinfo {author} {\bibfnamefont {S.}~\bibnamefont {Fan}},\ and\ \bibinfo {author} {\bibfnamefont {Z.}~\bibnamefont {Yu}},\ }\bibfield  {title} {\bibinfo {title} {Analog of superradiant emission in thermal emitters},\ }\href {https://doi.org/10.1103/physrevb.92.024302} {\bibfield  {journal} {\bibinfo  {journal} {Phys. Rev. B}\ }\textbf {\bibinfo {volume} {92}},\ \bibinfo {pages} {024302} (\bibinfo {year} {2015})}\BibitemShut {NoStop}%
\end{thebibliography}%

\end{document}